\documentclass[12pt]{spieman}  % 12pt font required by SPIE;
\usepackage{amsmath,amsfonts,amssymb}
\usepackage{graphicx}
\usepackage{setspace}
\usepackage{tocloft}
\usepackage{enumerate}

\usepackage{lineno}
\usepackage{comment}
\usepackage{slashbox}
\usepackage{subcaption}
\usepackage[export]{adjustbox}

\title{Adaptive optics with programmable Fourier-based wavefront sensors: a spatial light modulator approach to the LOOPS testbed}

\author[a,b,*]{Pierre Janin-Potiron}
\author[a,b]{Vincent Chambouleyron}
\author[c,b]{Lauren Schatz}
\author[b]{Olivier Fauvarque}
\author[d]{Charlotte Z. Bond}
\author[c]{Yannick Abautret}
\author[b]{Eduard Muslimov}
\author[b]{Kacem El-Hadi}
\author[a,b]{Jean-Fran\c cois Sauvage}
\author[b]{Kjetil Dohlen}
\author[b]{Beno\^it Neichel}
\author[b]{Carlos M. Correia}
\author[a,b]{Thierry Fusco}
\affil[a]{ONERA The French Aerospace Laboratory, F-92322 Ch\^atillon, France}
\affil[b]{Aix Marseille Univ, CNRS, CNES, LAM, Marseille, France}
\affil[c]{The University of Arizona, College of Optical Sciences, 1630 E University Blvd, Tucson, AZ 85719, USA}
\affil[d]{Institute for Astronomy
University of Hawaii-Manoa
640 N. Aohoku Place
Hilo, HI  96720 USA}

\cftpagenumbersoff{figure}
\cftpagenumbersoff{table} 
\begin{document} 
\maketitle

%%%%%%%%%%%%%%%%%%%%%%%%%%%%%%%%%%%%%%%%%
% ABSTRACT
%%%%%%%%%%%%%%%%%%%%%%%%%%%%%%%%%%%%%%%%%
\begin{abstract}

Wavefront sensors encode phase information of an incoming wavefront into an intensity pattern that can be measured on a camera.
Several kinds of wavefront sensors (WFS) are used in astronomical adaptive optics. Amongst them, Fourier-based wavefront sensors perform a filtering operation on the wavefront in the focal plane. The most well known example of a WFS of this kind is the Zernike wavefront sensor, and the pyramid wavefront sensor (PWFS) also belongs to this class. Based on this same principle, new WFSs can be proposed such as the n-faced pyramid (which ultimately becomes an axicone) or the flattened pyramid, depending on whether the image formation is incoherent or coherent.

In order to test such novel concepts, the LOOPS adaptive optics testbed hosted at the Laboratoire d'Astrophysique de Marseille has been upgraded by adding a Spatial Light Modulator (SLM). This device, placed in a focal plane produces high-definition phase masks that mimic otherwise bulk optic devices.

In this paper, we first present the optical design and upgrades made to the experimental setup of the LOOPS bench. Then, we focus on the generation of the phase masks with the SLM and the implications of having such a device in a focal plane. Finally, we present the first closed-loop results in either static or dynamic mode with different WFS applied on the SLM.
\end{abstract}

% Include a list of up to six keywords after the abstract

% Include email contact information for corresponding author
{\noindent \footnotesize\textbf{*}Pierre Janin-Potiron: \linkable{pierre.janin-potiron@lam.fr} }

\begin{spacing}{1}   % use double spacing for rest of manuscript

%%%%%%%%%%%%%%%%%%%%%%%%%%%%%%%%%%%%%%%%%
% INTRODUCTION
%%%%%%%%%%%%%%%%%%%%%%%%%%%%%%%%%%%%%%%%%
\section{Introduction}

Adaptive optics systems have been used in astronomy since the early 90s\cite{Rousset90,Rigaut91} in order to correct for atmospheric turbulence which degrades the quality of the observations from ground-based telescopes\cite{Roddier99}.
These systems were first deployed with Shack-Hartman wavefront sensors (SHWFS). In the late 90s, R. Ragazzoni proposed the pyramid wavefront sensor (PWFS\cite{Ragazzoni96}) which promised better performance in terms of sensitivity\cite{Ragazzoni99}, noise propagation\cite{Plantet15} and aliasing.

Within the last years, PWFS started being integrated on sky with promising results\cite{Esposito11}. There is still an effort to overcome the remaining challenges, such as handling the issue of optical gains\cite{Korkiakoski08,Deo18} or the non-linear behaviour of the sensor itself. To solve these issues, proposing and developing new and more robust wavefront sensing methods is necessary.
Based on the focal plane filtering principle, the PWFS is part of a WFS class called the Fourier-based WFS. We recently proposed to push further the  analytical description\cite{Verinaud04} of this class of WFS, with the introduction of a general formalize to describe all Fourier-based WFS\cite{Fauvarque17}.

Along with numerical and analytical studies, the \textit{Laboratoire d'Astrophysique de Marseille} (LAM) and the \textit{Office National d'\'Etudes et de Recherches en A\'erospatiales} (ONERA) developed an adaptive optics test bench for investigation on the PWFS\cite{Elhadi14,Bond15}. The LAM/ONERA On-sky Pyramid Sensor (LOOPS) optical bench is equipped with a 4 faces glass pyramid, a Shack-Hartman WFS, a tip-tilt modulation mirror and a $9\times9$ deformable mirror (DM). We present in this paper the upgrades we implemented on the LOOPS bench to provide a versatile tool for research on Fourier-based WFS.
Including the addition of a Spatial Light Modulator (SLM) in a focal plane, we created a configurable phase masks for the Fourier-based WFS class.

\begin{figure}[b!]
\includegraphics[width=.95\linewidth]{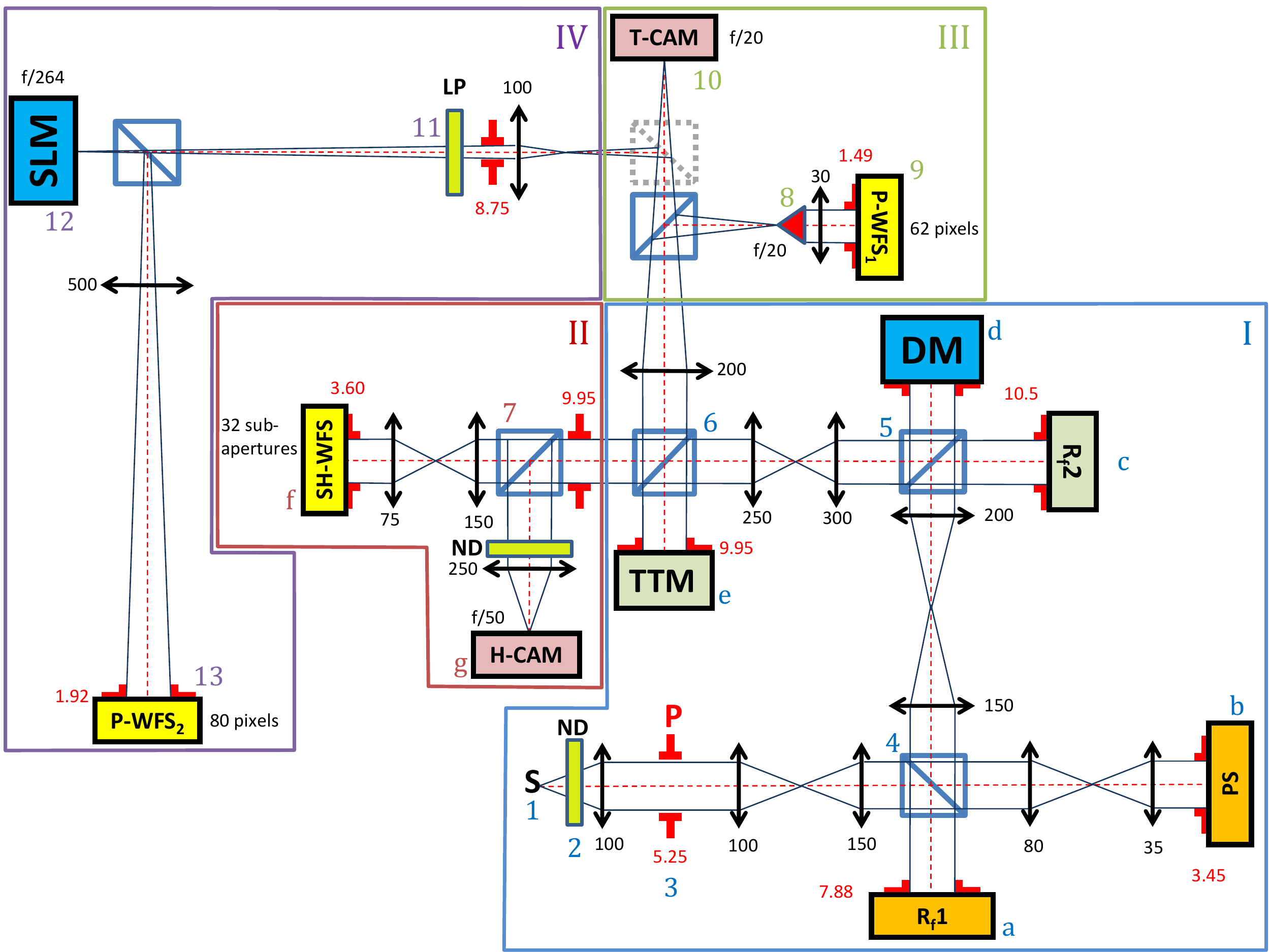}
\caption{Schematic view of the LOOPS bench. Each pupil plane is marked with a red aperture and its physical diameter is given. The f\# at the two cameras, pyramid and SLM positions are given as well.}
\label{LoopsScheme}
\end{figure}

We present in Sec.\ref{sec:LOOPSBench} the LOOPS upgraded optical design. The major components are detailed in order following the light path.
In Sec.\ref{sec:SLM}, we focus on the operation of the SLM and the care we took in order to produce the best possible images. We tackle challenges that come with operating an SLM including polarization, diffraction and phase wrapping effects, and propose a way to limit their impact on the WFS signal.
Then, we present in Sec.\ref{sec:firstimages} the first images acquired on the bench with different Fourier-based WFSs. We show as well the calibration process of the system using the Zernike modes and phase reconstruction.
In Sec.\ref{sec:cl4PWFS}, we finally present the results we obtained in closed-loop with a 4-faced PWFS applied onto the SLM using the slopes maps computational approach. We compare the residual wavefront error experimentally measured to the one obtained in a simulation of the bench and show a very good agreement.
We finally present the conclusions and expose the work in progress in Sec.\ref{sec:firstimages}.

%%%%%%%%%%%%%%%%%%%%%%%%%%%%%%%%%%%%%%%%%
% EXPERIMENTAL SETUP
%%%%%%%%%%%%%%%%%%%%%%%%%%%%%%%%%%%%%%%%%
\section{Experimental set-up: the LOOPS bench}
\label{sec:LOOPSBench}

Figure~\ref{LoopsScheme} shows the optical scheme of the bench. The principal components are presented in the list below. A complete description of the part I, II and III (see Fig.~\ref{LoopsScheme}) is available in a previous publication\cite{Bond15}.
The useful focal planes are pointed out with a black font value of the f\# at their respective locations. The pupil planes are marked with red apertures and their corresponding diameter is given with a red font.
The entrance pupil is defined by the deformable mirror, and the values correspond to the size of the re-imaged pupil along the path.
The cubes along the path are either cube beam-splitters, half-mirrors or flip-flop mirrors.

The bench is divided into four distinct blocks: (I) a common path (circled in blue on Fig.~\ref{LoopsScheme}), (II) a metrology path (circled in red on Fig.~\ref{LoopsScheme}), (III) a classic pyramid wavefront sensor path (circled in green on Fig.~\ref{LoopsScheme}) and (IV) a Fourier-based wavefront sensor path (circled in purple on Fig.~\ref{LoopsScheme}).

In (I) we have:
\begin{itemize}
    \item A monochromatic light source $[S]$ coming from a laser diode going through an optical fiber. The wavelength is $\lambda = 635$~nm.
    \item A reflective phase screen [PS] reproducing ground layer turbulence. This mask is mounted on a rotating plate and emulates the effect of a turbulent atmosphere. The strength of the turbulence is given by $d/r_0 = 3.2$ where $d$ is the actuator spacing and $r_0$ is the Fried parameter. This parameter was measured by fitting the mean variance of the first 100 Zernike coefficients measured at different locations on the phase screen.

    \item A continuous-sheet DM from ALPAO$^{\mbox{\scriptsize{\copyright}}}$  with 69 actuators. Nine actuators fit across the diameter of the pupil.
    The mirror has been calibrated to be used with either zonal control (i.e. actuator by actuator) or modal control (i.e. Zernike or Fourier modes).
    
    \item A PI$^{\mbox{\scriptsize{\copyright}}}$ piezoelectric tip-tilt mirror [TTM] that produces modulation (beam wander) at the pyramid apex. The modulation unit is operated at $500$ Hz and controlled in an open-loop mode (i.e. without feedback control on its position). A function generator produces the signal to control the mirror position and can generate an arbitrary waveform.
When running at $500$~Hz, the mirror can produce circular modulations with radii up to $10\lambda/D$.
Operated at lower frequency, the tip-tilt mirror can produce arbitrary modulation paths.

\end{itemize}

In (II) we have:
\begin{itemize}
    \item A Shack-Hartman wavefront sensor, that samples the pupil diameter with $32$ sub-apertures and provides reference wavefront measurements. The Shack-Hartman has been used to calibrate the Zernike and Fourier modes on the DM, and to measure the influence function of each of the actuators.
    \item A Hamamatsu$^{\mbox{\scriptsize{\copyright}}}$ ORCA-Flash CMOS camera with $1024 \times 1024$ pixels serving as the scoring camera to provide PSF and Strehl ratio measurements to assess the performance of the adaptive optics system.
    The Strehl ratio is $74\%$ for the best PSF and $3\%$ for the image obtained with the turbulent phase screen.
\end{itemize}

In (III) we have:
\begin{itemize}
    \item  monolithic	4-sided glass pyramid.
\item an OCAM$^2$ EM-CCD camera [P-WFS$_1$] with $240 \times 240$ pixels. The pupil is sampled by 62 pixels.
\item a Thorlabs camera [T-CAM] used to monitor the modulation path delivered by the tip-tilt modulation mirror.
\end{itemize}

An finally in (IV):
\begin{itemize}
    \item A Hamamatsu Spatial Light Modulator (LCOS-SLM X13138) with $1024 \times 1280$ pixels that produces arbitrary phase screens in the focal plane. The use of this device, initially proposed by Akondi \textit{et al}.\cite{Akondi13} and applied on a real testbed more recently by Engler \textit{et al}.\cite{Engler18}, promises a reliable modular phase screen generator. We go more into details on the SLM operation and control in Section~\ref{sec:SLM}.
\item a second OCAM$^2$ EM-CCD camera [P-WFS$_2$], identical to the one used on the classic pyramid path. The pupil is sampled by 80 pixels.
\end{itemize}

The LOOPS bench is operated using the Object Oriented Matlab Adaptive Optics (OOMAO\cite{Conan14}) toolbox. The maximum closed-loop frequency is 300~Hz.

%%%%%%%%%%%%%%%%%%%%%%%%%%%%%%%%%%%%%%%%%
% SLM
%%%%%%%%%%%%%%%%%%%%%%%%%%%%%%%%%%%%%%%%%
\section{Phase masks generation with a spatial light modulator}
\label{sec:SLM}

The spatial light modulator uses liquid crystal technology and exploits the birefringence properties of the crystals to produce easily configurable phase masks. The crystals are oriented using an adjustable electric field which allows to chose the phase delay pixel by pixel, thus creating the desired phase map.
Limitations are due to the discrete nature of a pixelated phase mask, as well as the fact that the light has to be linearly polarized to match the orientation of the crystals.
WFS creation with a SLM have already been demonstrated in Engler \textit{et al}.\cite{Engler18}. In this paper, authors presented a method to produce Fourier-based wavefront sensor around the $0^{th}$ diffraction order (see bellow) which might, in some situations, interfere with the signal of interest.
We describe in this section the operating method we followed to produce the best possible images on [P-WFS$_2$] after the SLM.

\begin{figure}[tb!]
\begin{center}
\includegraphics[width=.3\linewidth]{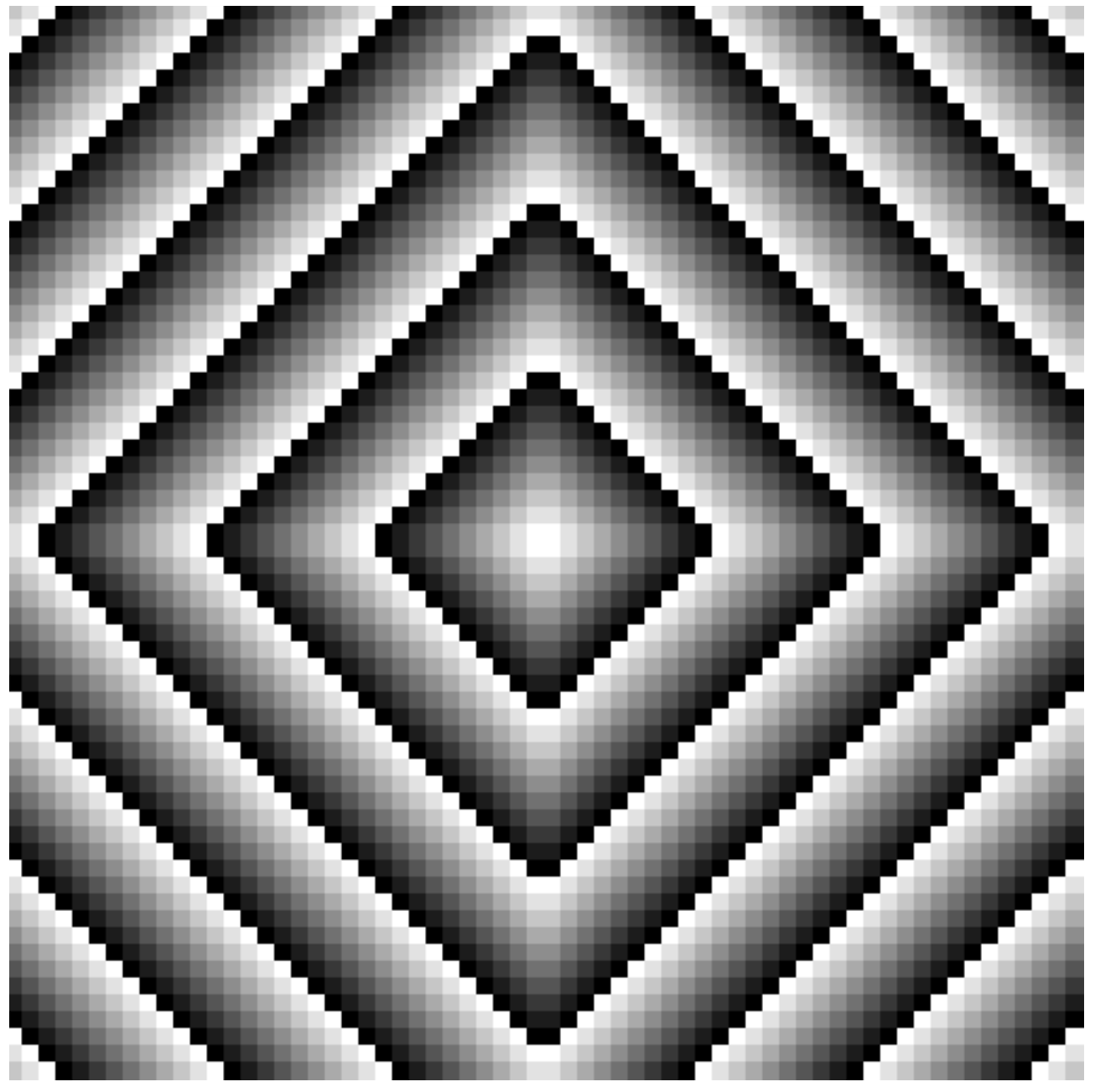}
\hfill
\includegraphics[width=.3\linewidth]{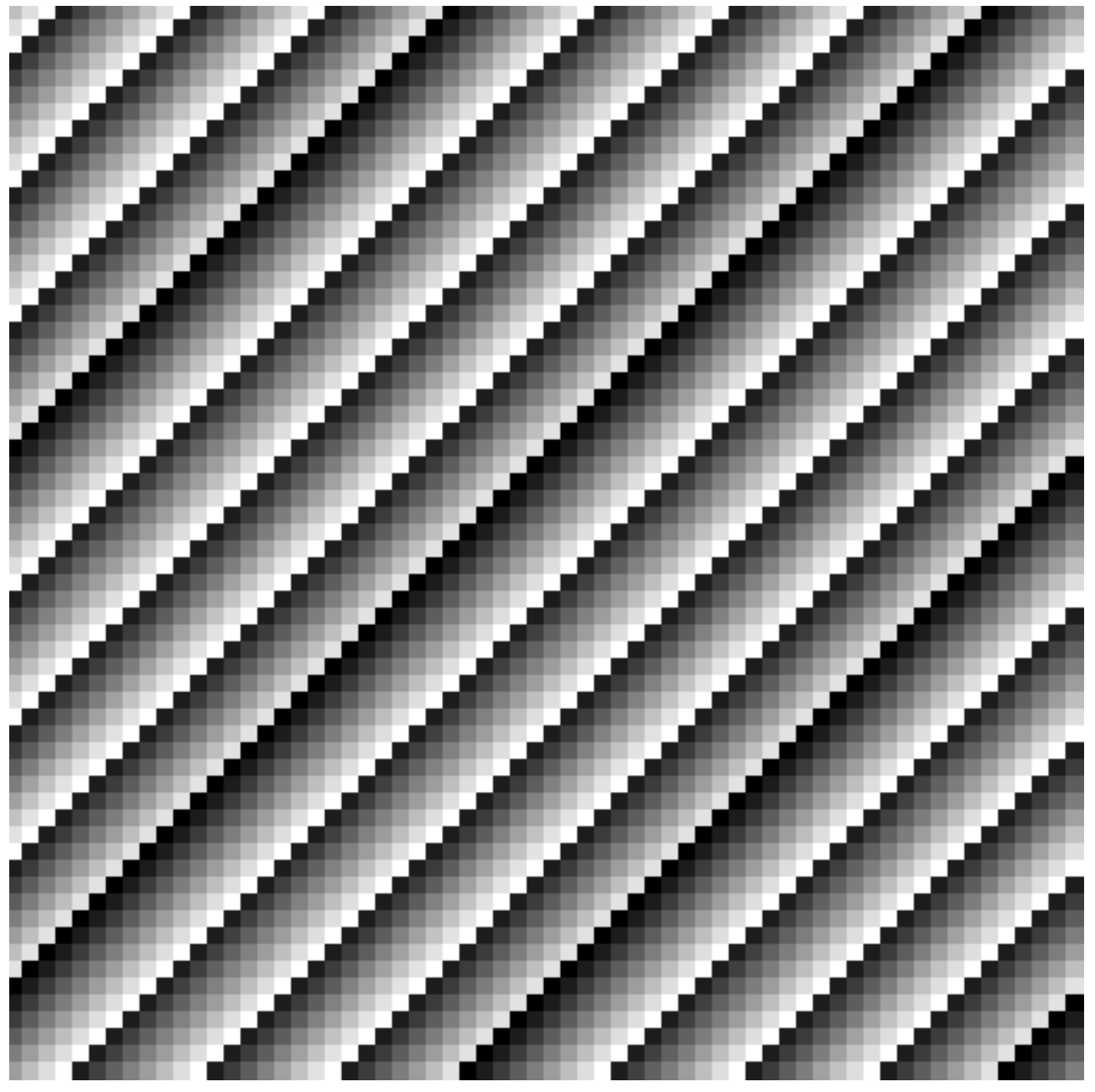}
\hfill
\includegraphics[width=.3\linewidth]{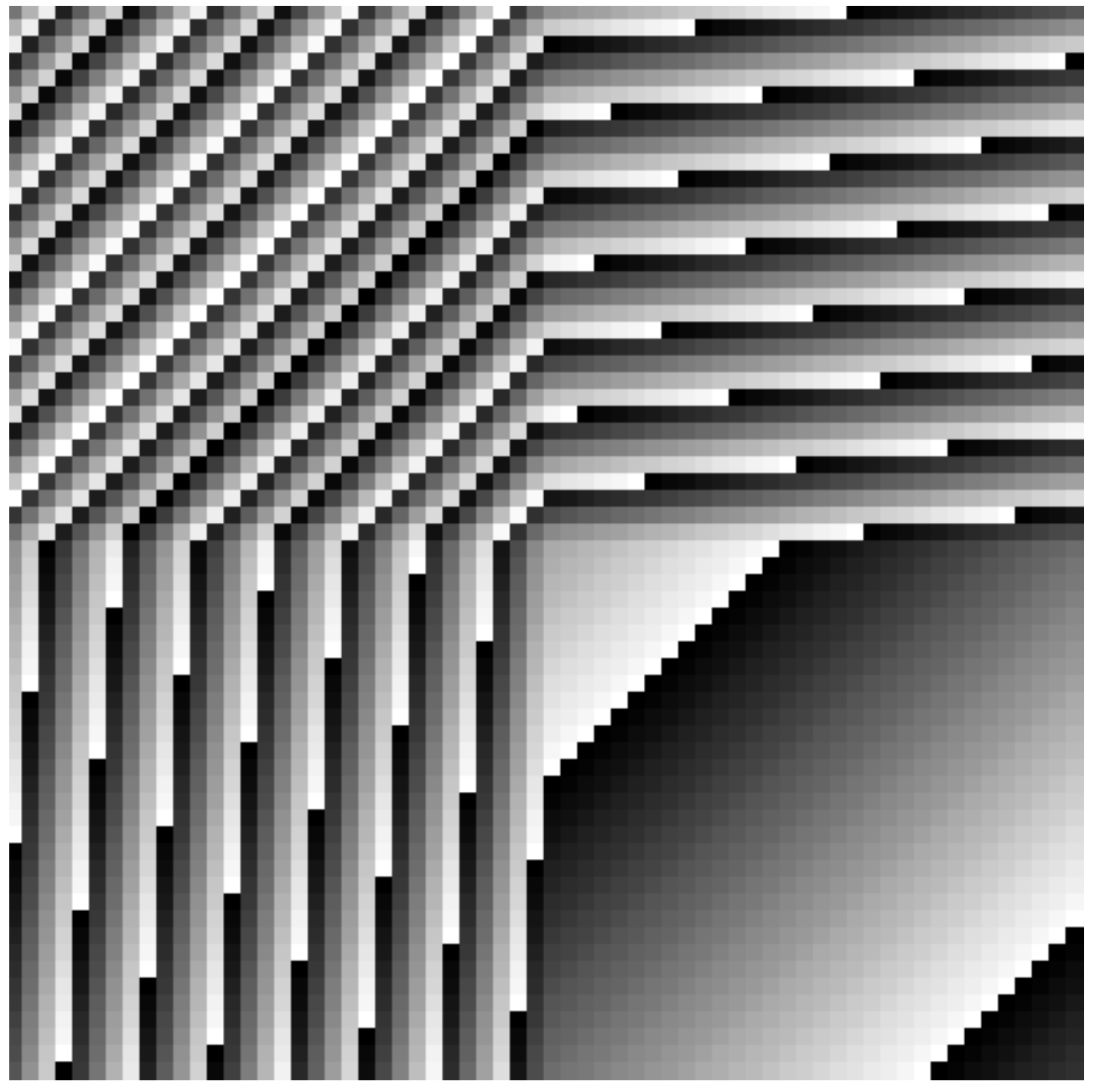}

\begin{subfigure}{.3\textwidth}
\centering
\includegraphics[width=\linewidth]{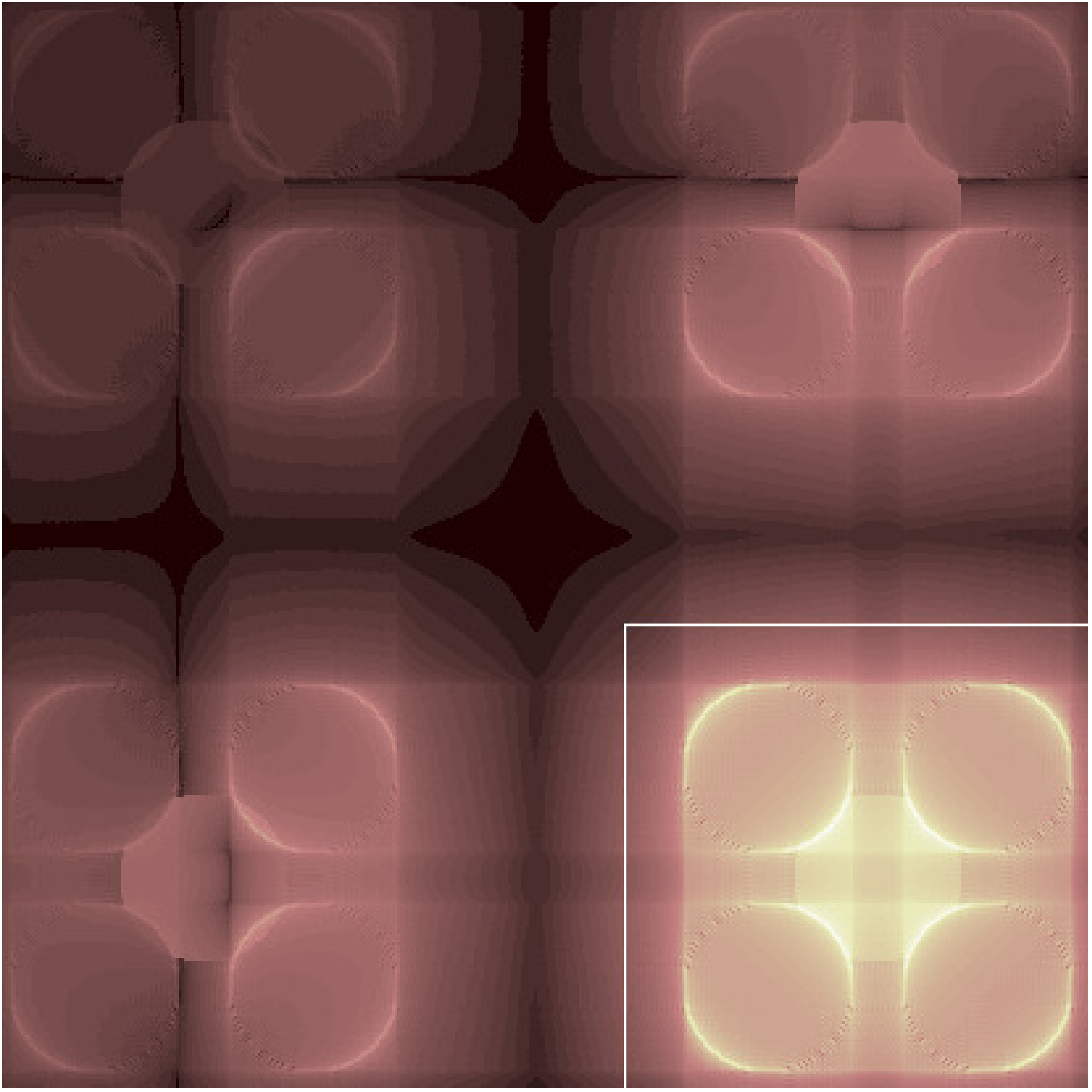}
\caption{}
\label{sub:bitmapA}
\end{subfigure}
\hfill
\begin{subfigure}{.3\textwidth}
\centering
\includegraphics[width=\linewidth]{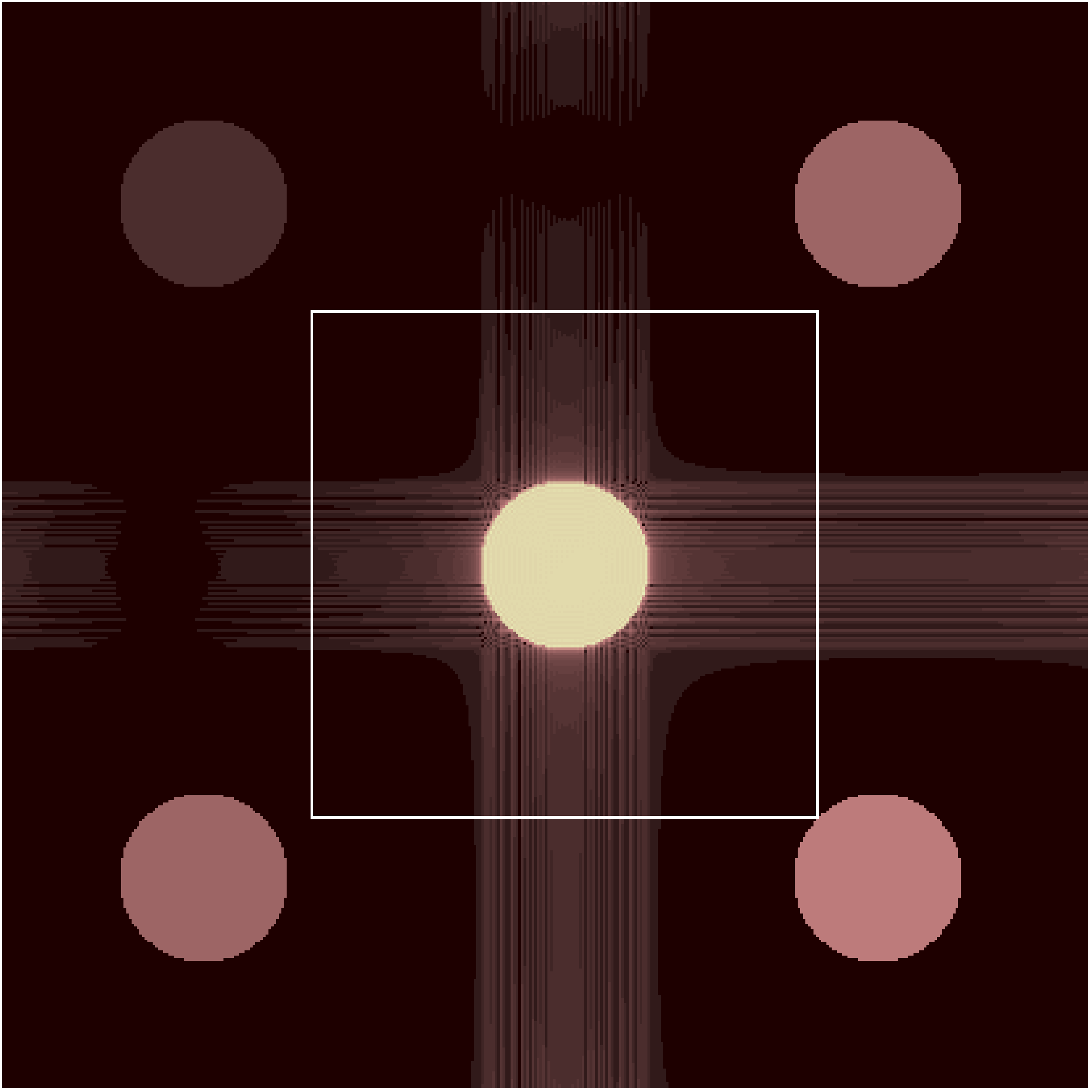}
\caption{}
\label{sub:bitmapB}
\end{subfigure}
\hfill
\begin{subfigure}{.3\textwidth}
\centering
\includegraphics[width=\linewidth]{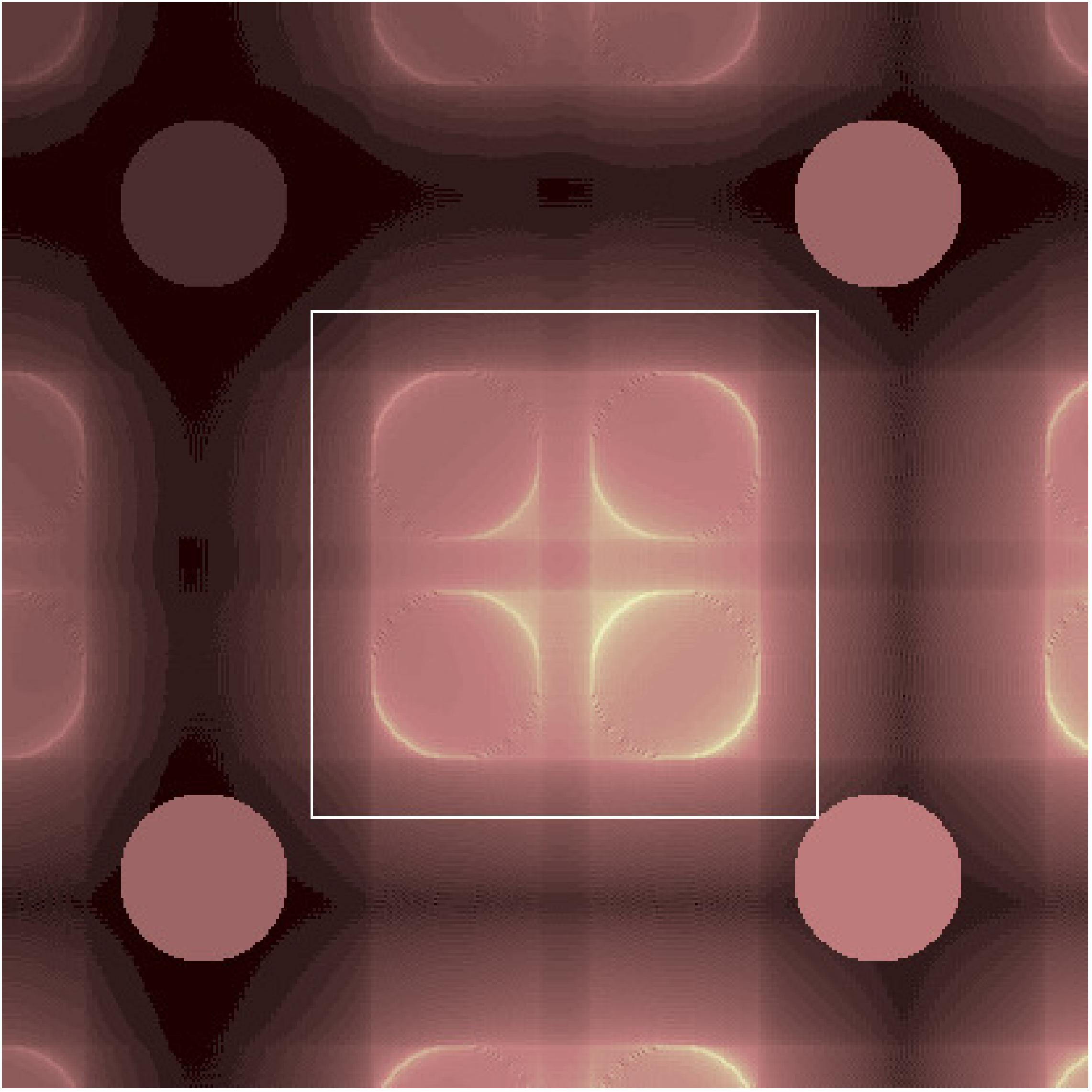}
\caption{}
\label{sub:bitmapC}
\end{subfigure}

\caption{Top: Bitmaps applied on the SLM to produce (a) a centered 4 faced pyramid, (b) a shifted pupil and (c) a shifted 4 faced pyramid. Bottom: corresponding pupil plane images obtained with these masks. The white square represents the OCAM imprints. The pixel values are set between 0 and 220, the value corresponding to $2\pi$ at the wavelength of operation.}
\label{fig:SLMwrap}
\end{center}
\end{figure}

\begin{itemize}

\item \textbf{Polarization}

As stated previously, the polarization of the light incident on the SLM has to be linear and orientated in the direction of the crystals.
Let us assume that the crystals are vertically oriented. The light after the polarizer will inevitably produce both a horizontal and a vertical component. This is because the polarizer is never perfectly aligned, and the extinction ratio between the vertical and horizontal component of a real polarizer is never exactly null.
The horizontally polarized light not being affected by the SLM, produces an unmodified pupil image at the exact same position in the pupil plane as if the SLM was a mirror, with a flux depending on the efficiency of the polarization.
On the LOOPS bench, the polarizer has an extinction ratio of $700:1$. It means there is at least $1/700$ of the total flux going back into the $0^{th}$ diffraction order.

\item \textbf{Filling factor}

Gaps between the pixels of the SLM are also present. The light reflecting on the SLM at these positions is therefore not affected by phase modulation.
The fill factor of the SLM used on the LOOPS bench is $96\%$, meaning that $4\%$ of the vertically polarized incoming flux is not modulated.
This contributes to create an unmodulated pupil image on the pupil plane. This means that even in the ideal scenario where we have a perfect polarizer, there will still be a pupil image gathering $4\%$ of the incoming flux located at the unmodulated pupil position. This contribution is vertically polarized and does not interfere with the contribution due to the extinction ratio or bad orientation of the polarizer which is horizontal.

\item \textbf{Diffraction}
The SLM operates by diffracting the light, similar in principle to a computer generated hologram (CGH\cite{Slinger05,Yang14}) or diffraction grating\cite{Born99}. One consequence is that the SLM produces multiple diffraction orders containing the same pupil image. Using geometric optics we can relate the positions of the diffraction patterns to the number of pixel of SLM $n_p$ per $\lambda/D$. This behaviour can be seen on Fig.~\ref{fig:SLMwrap} (bottom image a) where the central pattern is repeated on the sides of the image.
Calling $n_x$ and $n_y$ the diffraction orders, their location $P(n_x,n_y)$ in the pupil plane is given by
\begin{equation}
    P(n_x,n_y) = \sqrt{(n_x^2+n_y^2)} n_p D.
\end{equation}
The distribution of the diffraction orders will put constraints on the choice of $n_p$ so that there is no superimposition of the signals.
IN addition we also require avoidance of central diffraction order due to the imperfect polarization or the filling factor. This case can be seen on Fig.~\ref{fig:SLMwrap} (bottom image a), where the $0^{th}$ diffraction order degrades the signal from the pyramid. To correct this we apply an additional tip-tilt onto the SLM to move the central order off our detector.
Knowing the size of $l_d$ of the detector we put in the pupil plane (on LOOPS the detector is square), we can express the minimal tip-tilt $\gamma$, given in rad/px, to apply so that the detector does not see the $0^{th}$ order of diffraction:
\begin{equation}
\label{eq:gammalim}
    \gamma_{lim} \ge 2\pi \frac{l_d/D + \sqrt{2}/2}{n_p}.
\end{equation}
Figure~\ref{fig:SLMwrap} (bottom image b) shows the tilt at this limit applied on the SLM. The camera imprint is shown with the yellow square. We see that the bottom right hand corner of the detector is touching the central order. The combination of $\gamma$ and the phase to produce the pyramid is finally presented on Figure~\ref{fig:SLMwrap}  (c) where we see the pyramid signal being free of the central diffraction order.
Due to the non-unitary filling factor, there is a cardinal sine envelope modulating the intensity\cite{Born99} on the images obtained after the SLM. It means that the greater the tip-tilt added to the phase of the SLM the lower the intensity of the tilted pyramid signal. By increasing the number of pixel per diffraction element in the focal plane, this effect is attenuated. In the limit conditions at $\gamma_{lim}$ on the LOOPS bench, the intensity loss between the extremal point of a contiguous 4 faces pyramid is around $4\%$.

\begin{figure}[b!]
\begin{center}
\includegraphics[height=.332\linewidth,valign=t]{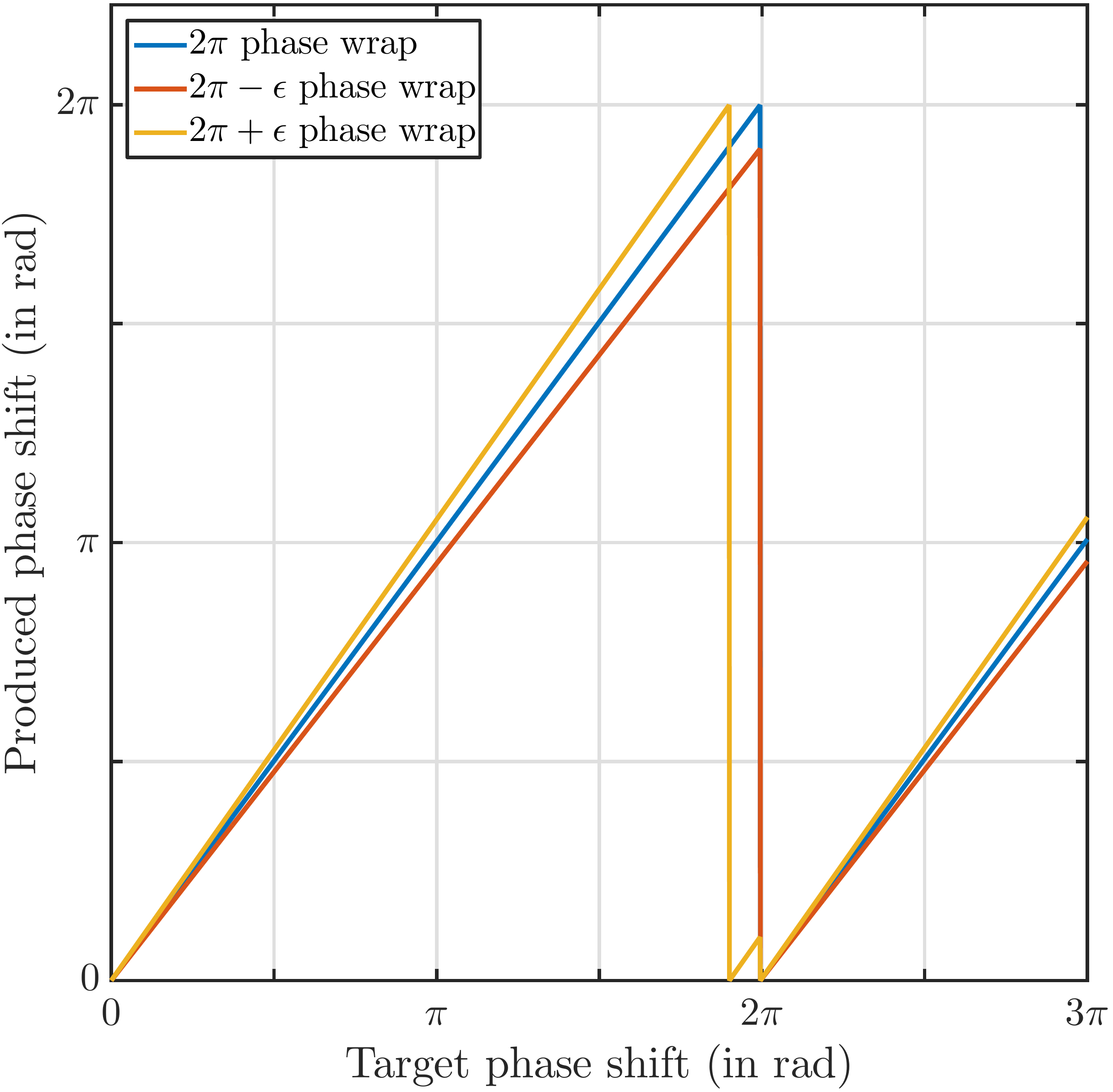}
\includegraphics[height=.3\linewidth,valign=t]{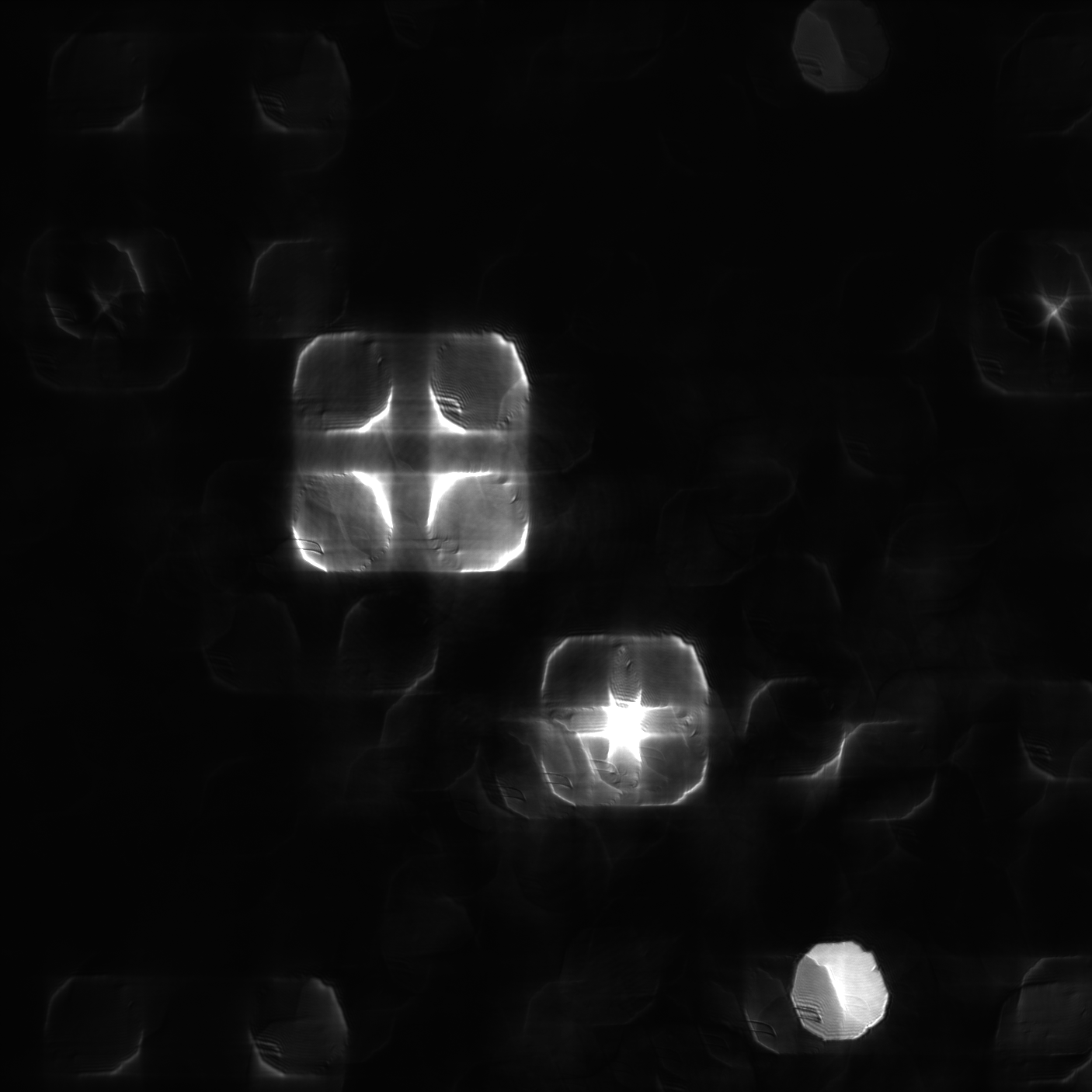}
\includegraphics[height=.3\linewidth,valign=t]{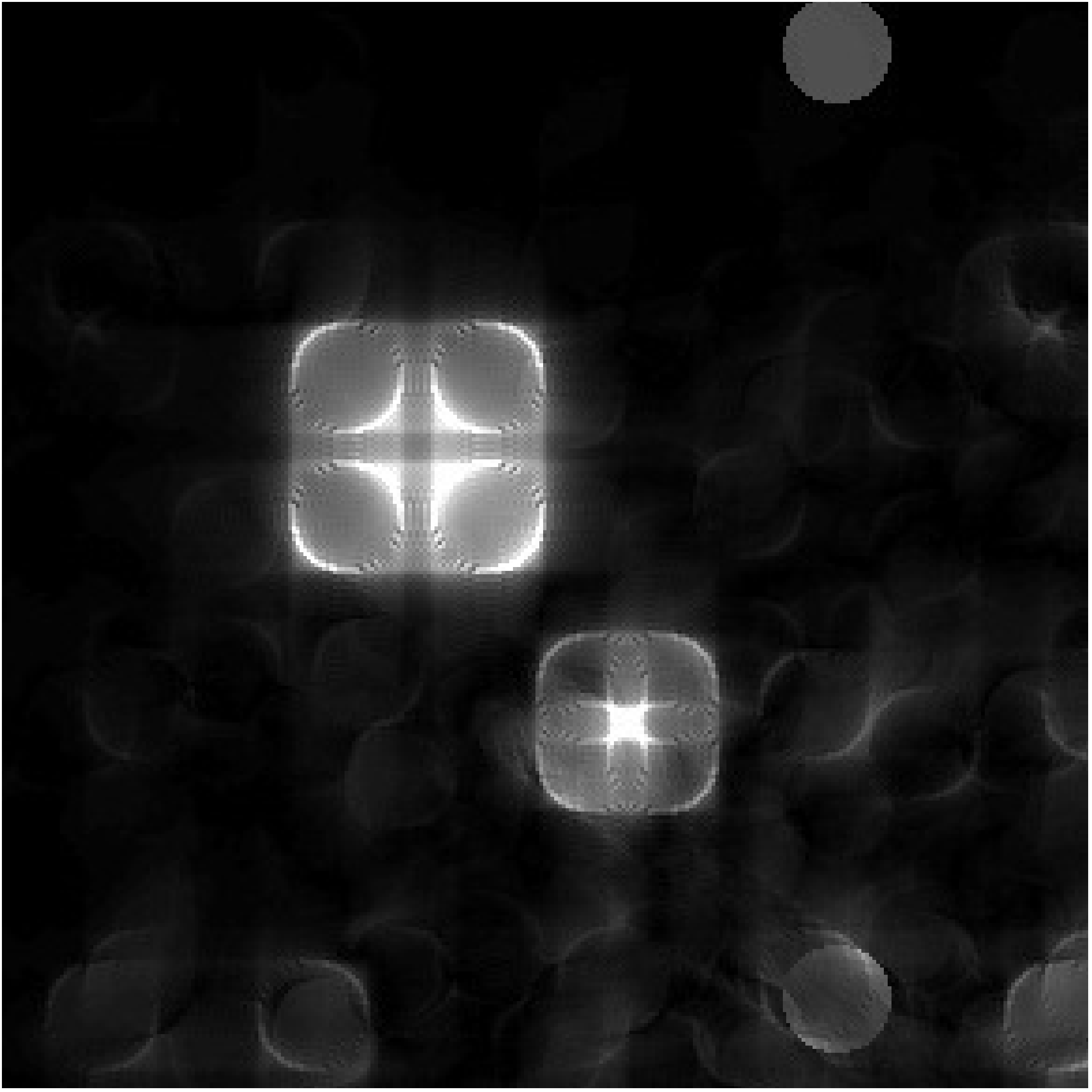}
\caption{Left: output phase displayed by the SLM vs. targeted phase in the case of a perfect $2\pi$ wrap (blue), a $2\pi-\epsilon$ wrap (red) and a $2\pi+\epsilon$ wrap (yellow). Center: saturated image obtained on a separated optical bench of a 4 faces PWFS in the $2\pi+\epsilon$ phase wrap case. Right: corresponding simulation image.} 
\label{fig:PhaseWrapping}
\end{center}
\end{figure}

Finally, for a given number $N$ of faces on our pyramid, the minimal gradient to apply to each faces in order to make sure the pupil are not superimposed is:
\begin{equation}
    \gamma_{pyr} \ge \frac{D}{2sin(\pi/N)}.
\end{equation}

\item \textbf{Phase wrapping}

The optical path difference produced by the SLM has a limited range. It means that, at a given wavelength, the phase retardation one can obtain is between $0$ and $2\pi + \delta_\Phi(\lambda)$. At the wavelength of the laser on the LOOPS bench ($\lambda=660$~nm), the phase retardation can be tuned between $0$ and $2.5\pi$ and the $0$ to $2\pi$ range corresponds to $220$ bits. If not properly calibrated, it is possible that the setup wraps the phase between $0$ and a value different from $2\pi$. Such configurations are presented on Fig.~\ref{fig:PhaseWrapping} (left), where we show a perfect $2\pi$ phase wrap (blue curve), a phase step lower than $2\pi$ (red curve), and a phase step greater than $2\pi$ which introduces small additional bumps on the phase (yellow curve). Fig.~\ref{fig:PhaseWrapping} (center) shows the effects of phase wrapping on the pupil plane. Fig.~\ref{fig:PhaseWrapping} (right) is an image in a numerical simulation of our testbed including the effects of phase wrapping.

\end{itemize}

\newcommand{\figsize}{0.24\linewidth}
\begin{figure}[b!]
\begin{center}

\includegraphics[width=\figsize]{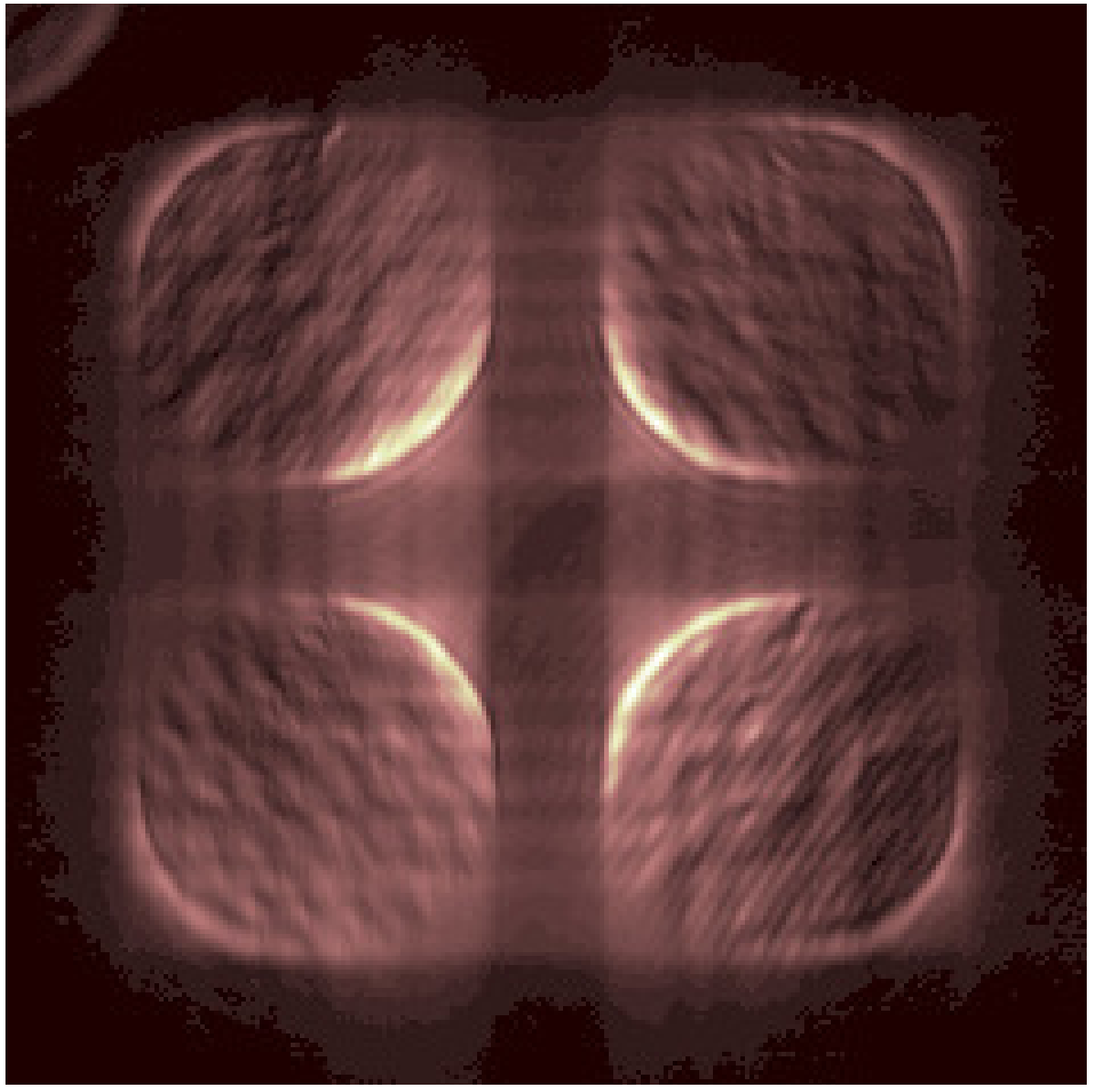}
\includegraphics[width=\figsize]{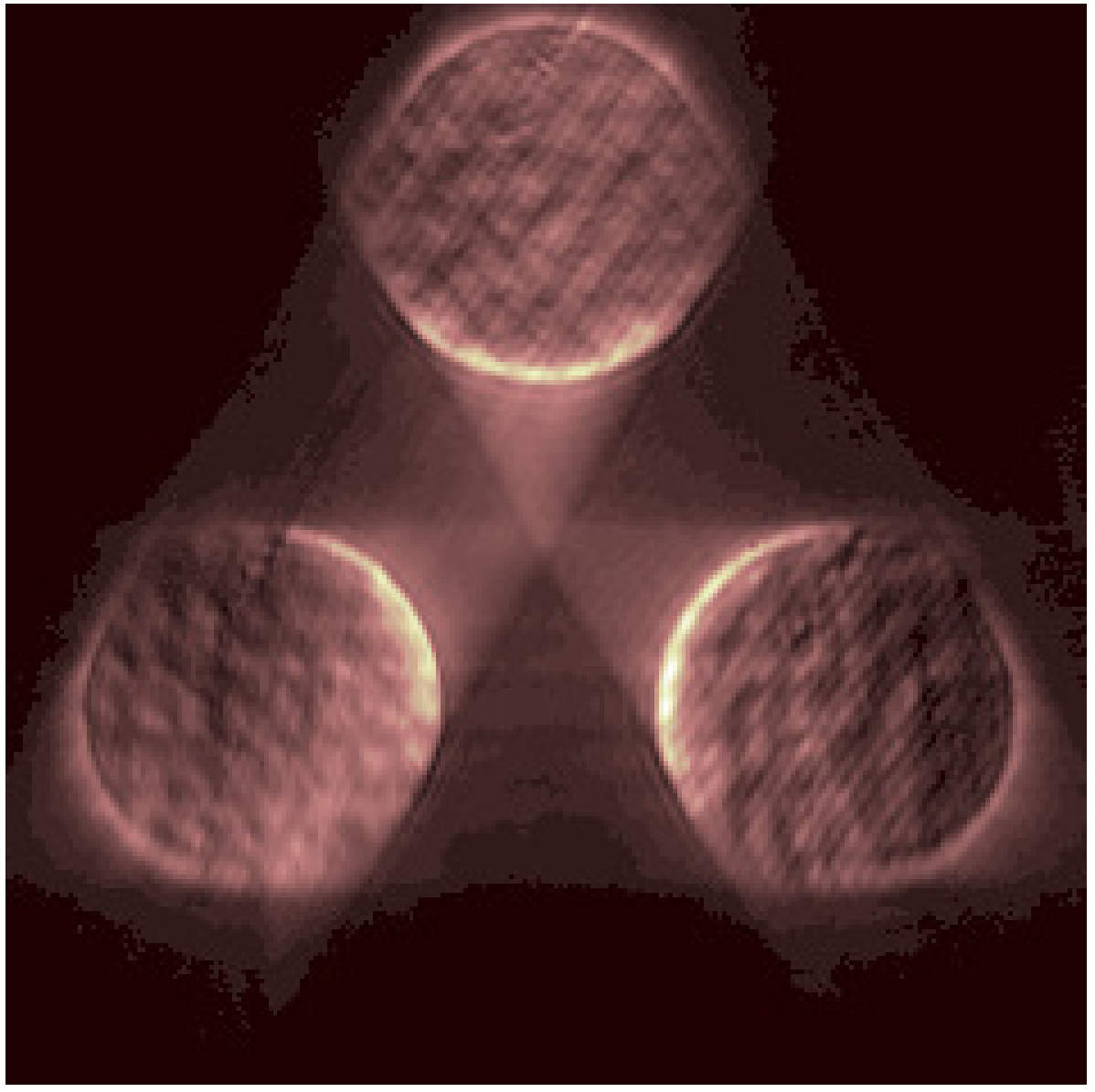}
\includegraphics[width=\figsize]{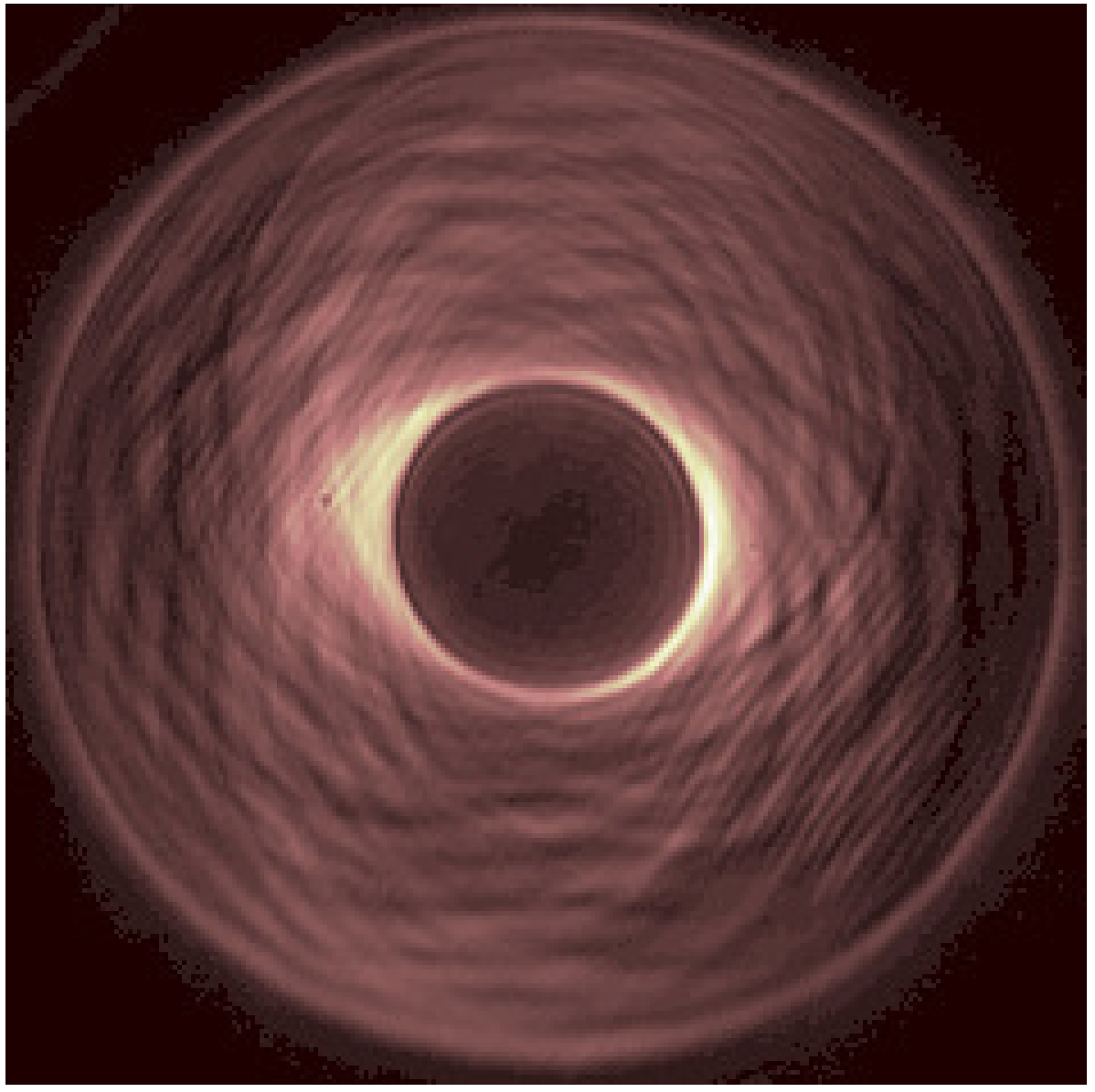}
\includegraphics[width=\figsize]{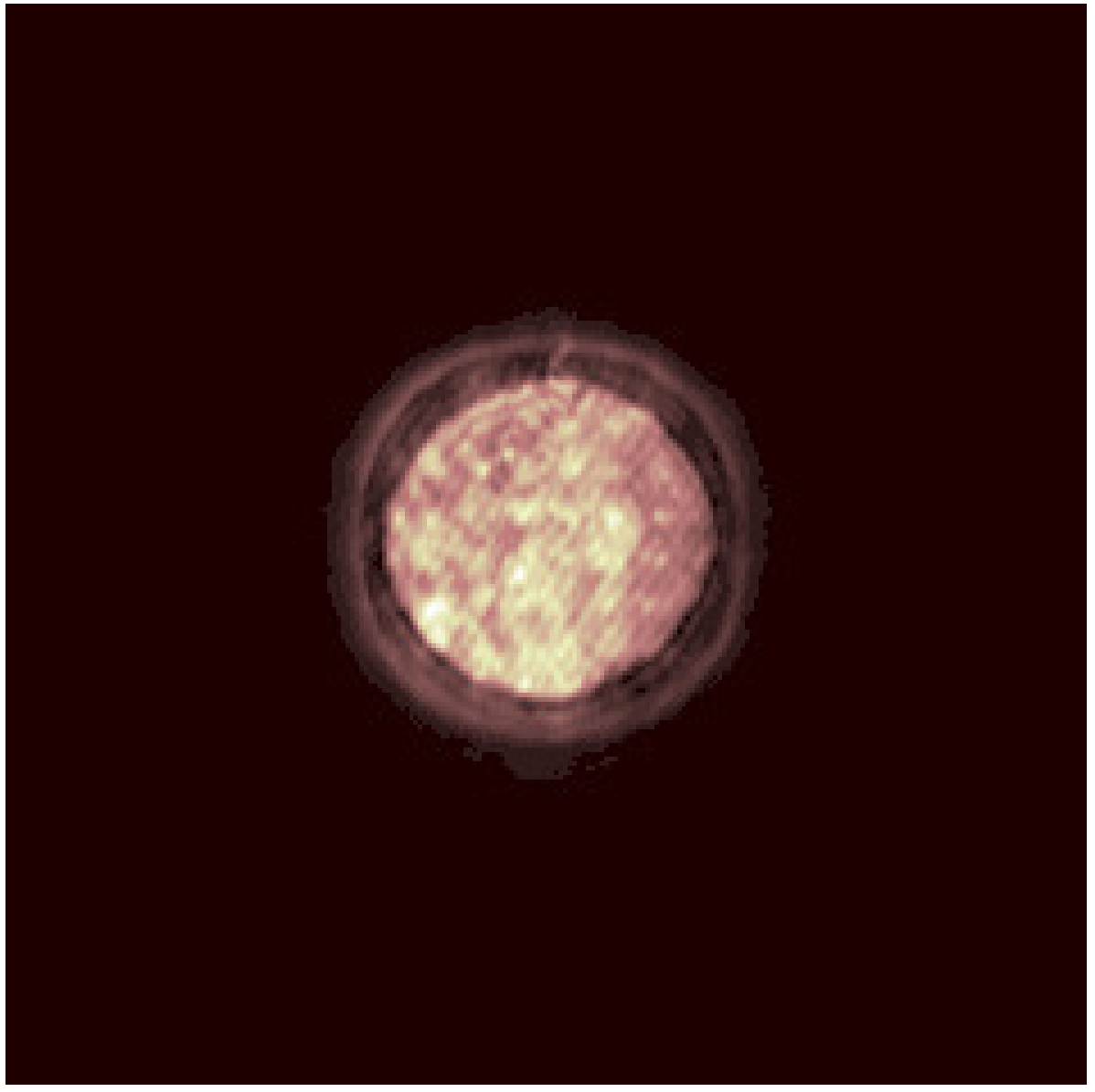}
\\
\begin{subfigure}{\figsize}
\centering
\includegraphics[width=\linewidth]{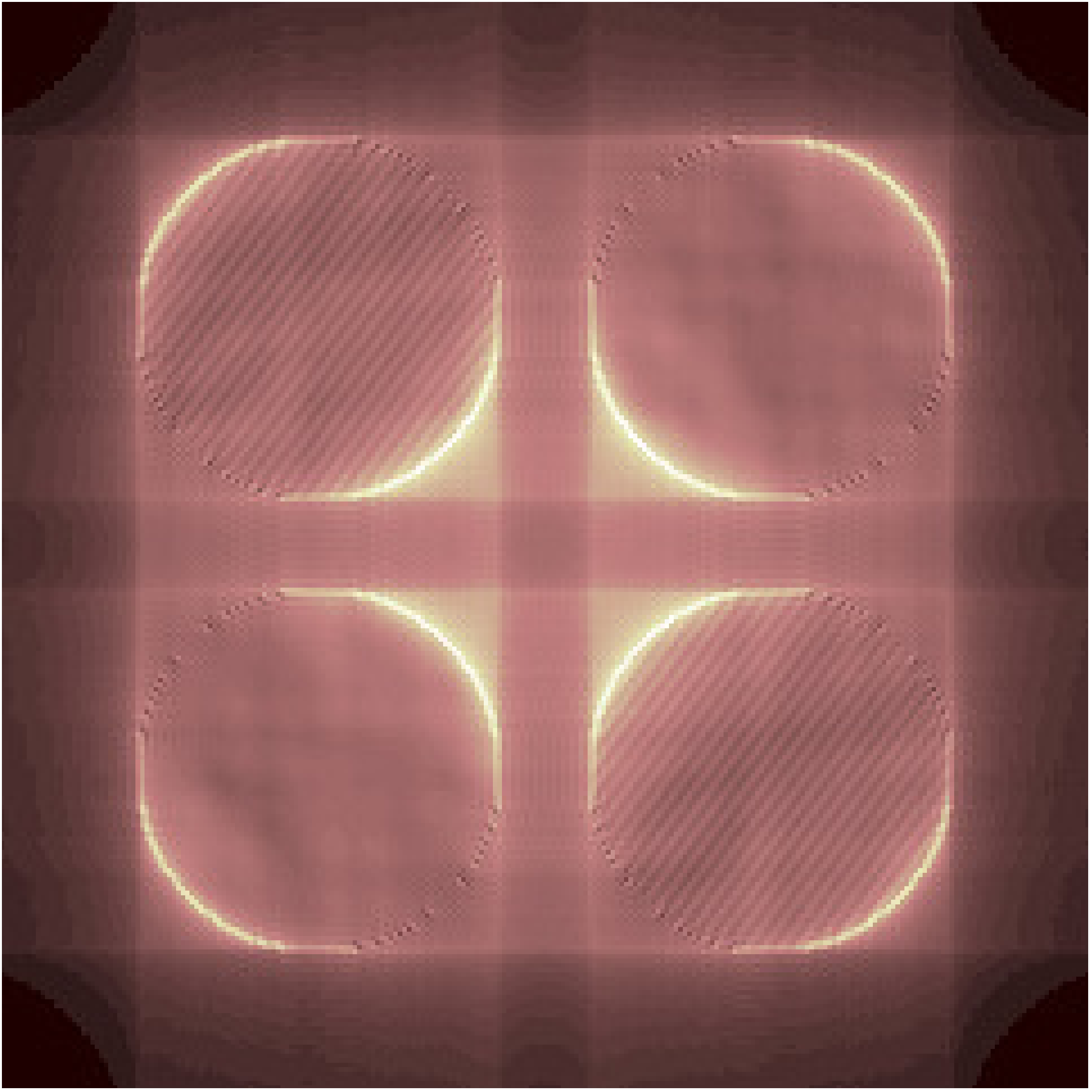}
\caption{}
\label{sub:unmoda}
\end{subfigure}
\begin{subfigure}{\figsize}
\centering
\includegraphics[width=\linewidth]{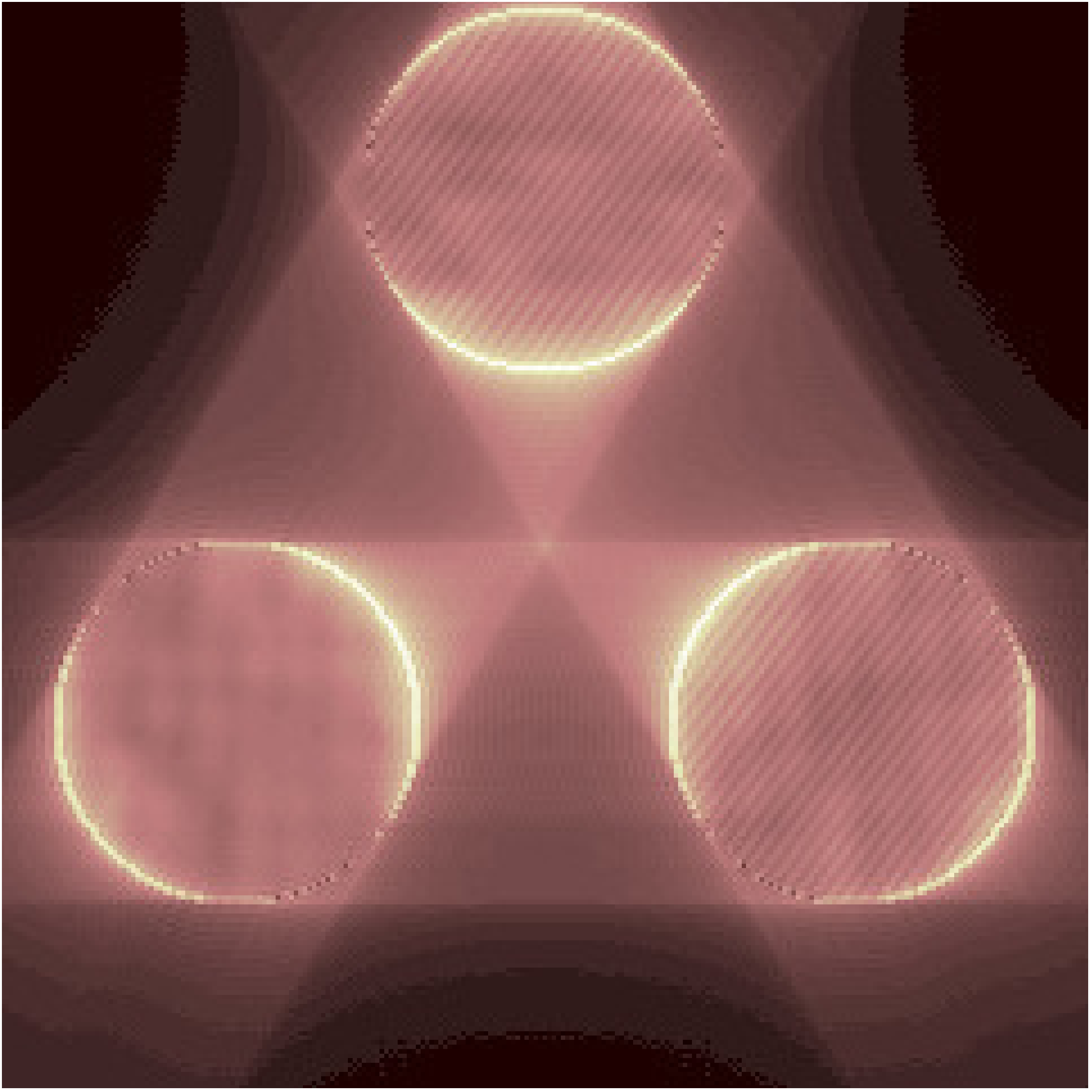}
\caption{}
\label{sub:unmodb}
\end{subfigure}
\begin{subfigure}{\figsize}
\centering
\includegraphics[width=\linewidth]{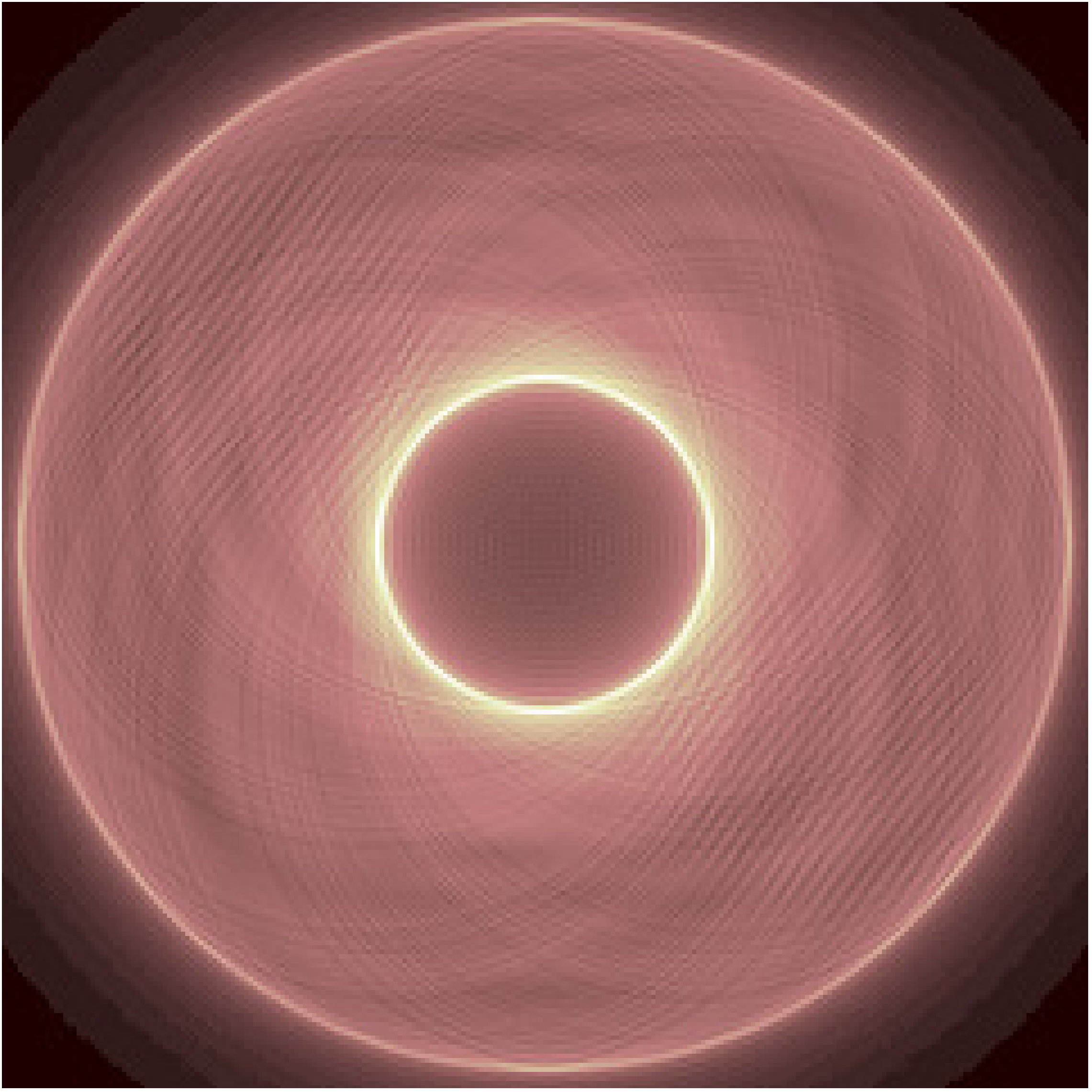}
\caption{}
\label{sub:unmodc}
\end{subfigure}
\begin{subfigure}{\figsize}
\centering
\includegraphics[width=\linewidth]{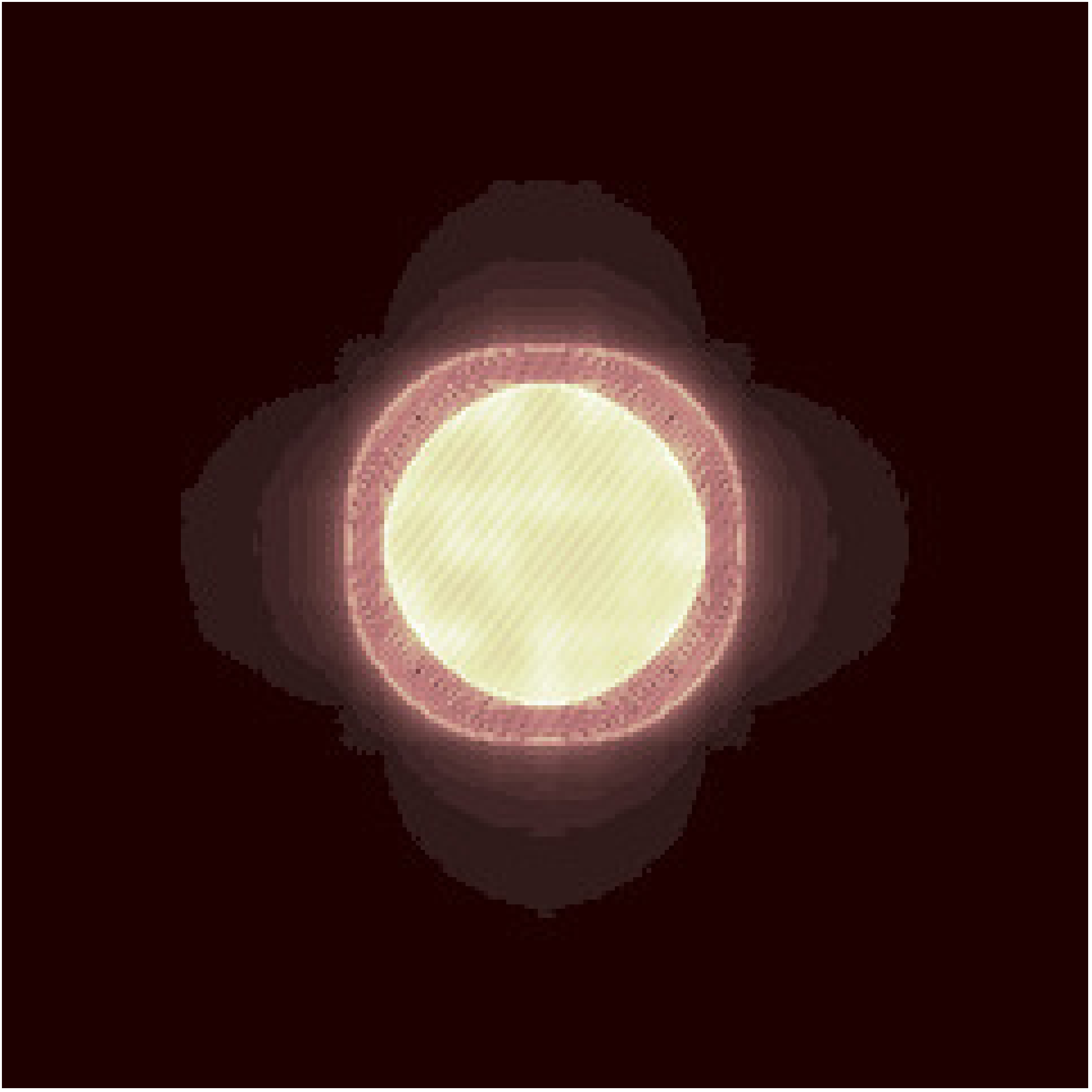}
\caption{}
\label{sub:unmodd}
\end{subfigure}
\caption{Top: images taken on the LOOPS bench in the pupil plane after the SLM. Bottom: simulation images corresponding to the top cases. The focal masks applied on the SLM are respectively: (a) a 4 faces pyramid as already presented in Fig~\ref{sub:bitmapC}, (b) a 3 faces pyramid as already presented, (c) an axicone and (d) a 4 faced flattened pyramid.}
\label{fig:SLMBench}
\end{center}
\end{figure}

We put all of the above into consideration when we created the different Fourier-based wavefront sensor phase masks. Our polarizer has a $700:1$ extinction ratio and our SLM a $96\%$ filling factor, which means there is around $4\%$ of the incoming light gathered on the $0^{th}$ diffraction order. We use an additional tilt slightly greater than $\gamma_{lim}$ presented in Eq.~\ref{eq:gammalim}. We present the results obtained on the LOOPS bench with this setup in Sec.~\ref{sec:firstimages}.

%%%%%%%%%%%%%%%%%%%%%%%%%%%%%%%%%%%%%%%%%
% STATIC EVALUTION
%%%%%%%%%%%%%%%%%%%%%%%%%%%%%%%%%%%%%%%%%
\section{Static evaluation}
\label{sec:firstimages}
Figure~\ref{fig:SLMBench} presents the images acquired on the LOOP bench (top) along with their respective simulations (bottom) for the 4 and 3 faced pyramids (images a and b), the axicone (image c) and the flattened pyramid (image d). There is a good agreement between the images produced in simulation and the images obtained on the bench which is promising in terms of quality in the phase measurement.
 The residuals that can be seen on the bench images are either residual phase on the wavefront (for example the recognizable waffle pattern coming from the DM itself) or phase aberrations in the focal plane, transforming into intensity variation in the pupil plane. In order to prevent this last point, we apply on the SLM the phase flatmap correction provided by Hamamatsu.
 Some fringes are present on the images as well. They come from interference between the WFS image and ghosts coming from the multiple cube beam splitters. The contrast of these fringes is around 5\%. These ghosts are too close to the PSF core to be filtered out optically. This pattern is static and disappears after the calibration stage.
 We added this effect to the simulated images on Fig.\ref{fig:SLMBench} by applying a fainter second source at the location of the brightest ghost we measured on the focal image at the SLM position.
 It could as well be numerically filter afterward as it is way beyond the control frequency defined by the actuators pitch.
 Manufacturing errors such as rooftop defects, difference in faces angles or asymmetry in the masks can be produced by the SLM. The relatively high sampling of the focal plane insures that the mask can be considered as continuous. The impact of such defects on the performances of the different WFSs can then be assessed and classified as critical for the operation or not.
The LOOPS bench thus offers the opportunity to test a large variety of configurations.

\renewcommand{\figsize}{0.25\linewidth}
\begin{figure}[b!]
\begin{center}
\includegraphics[width=\figsize]{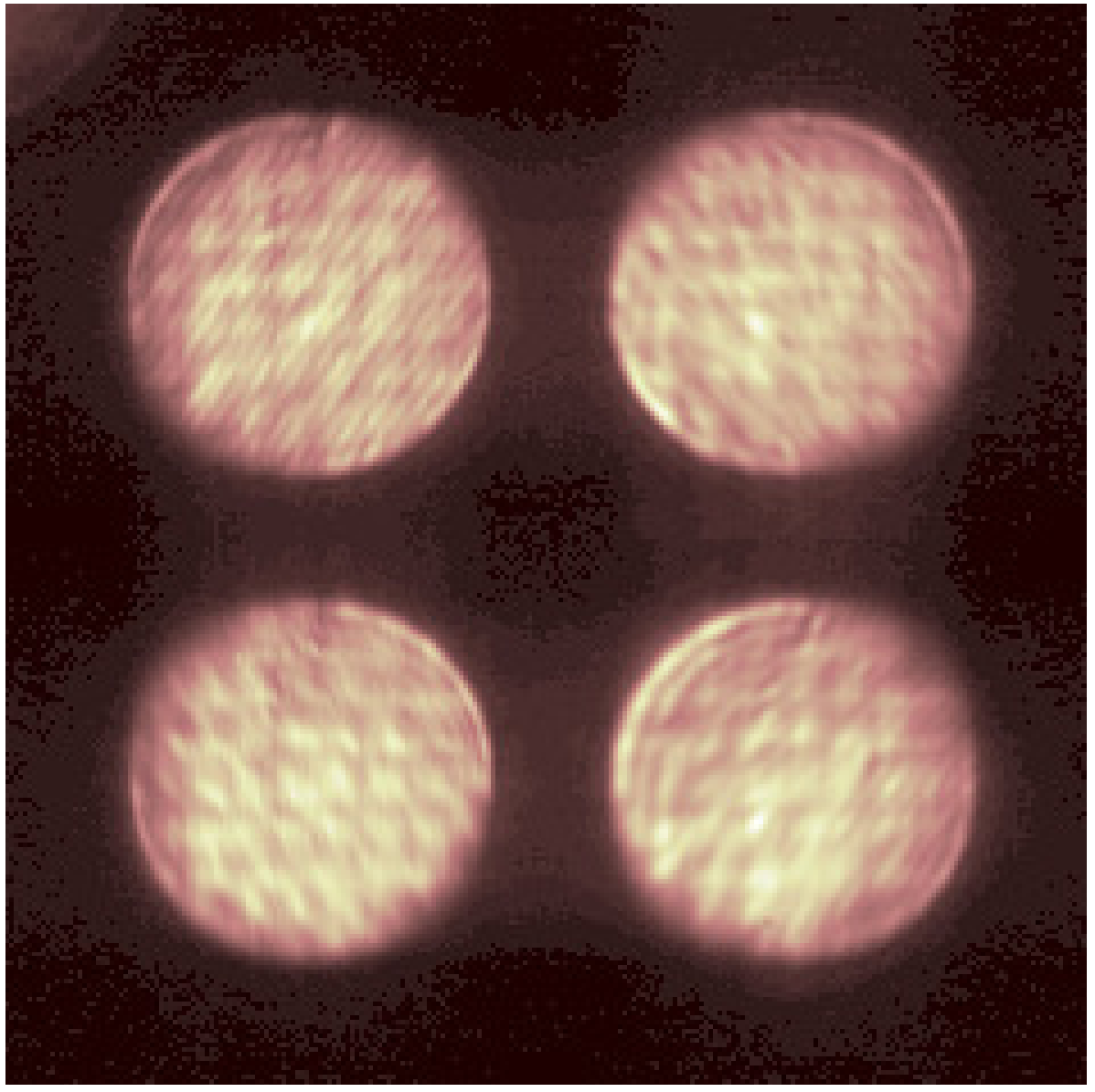}
\includegraphics[width=\figsize]{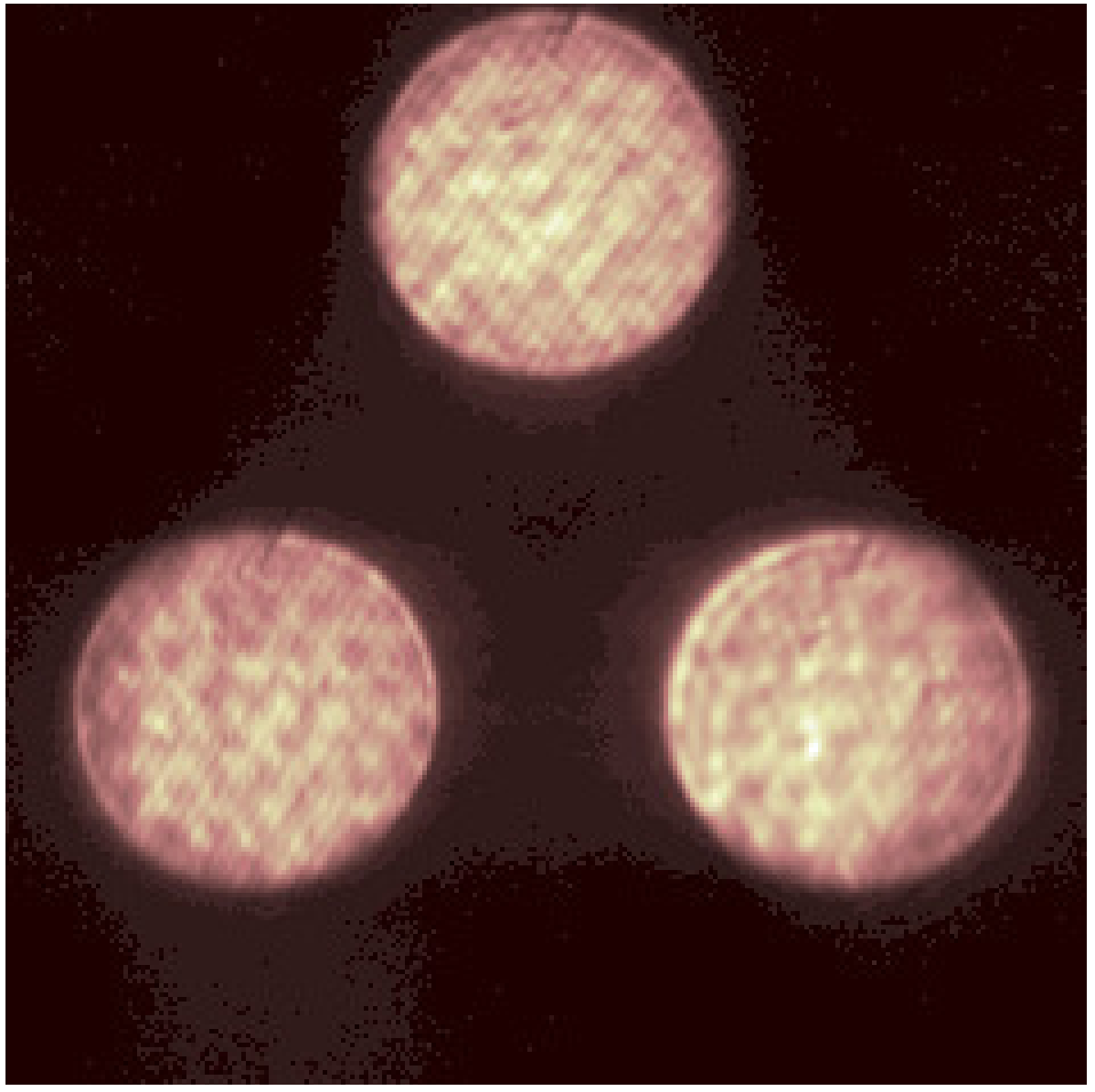}
\includegraphics[width=\figsize]{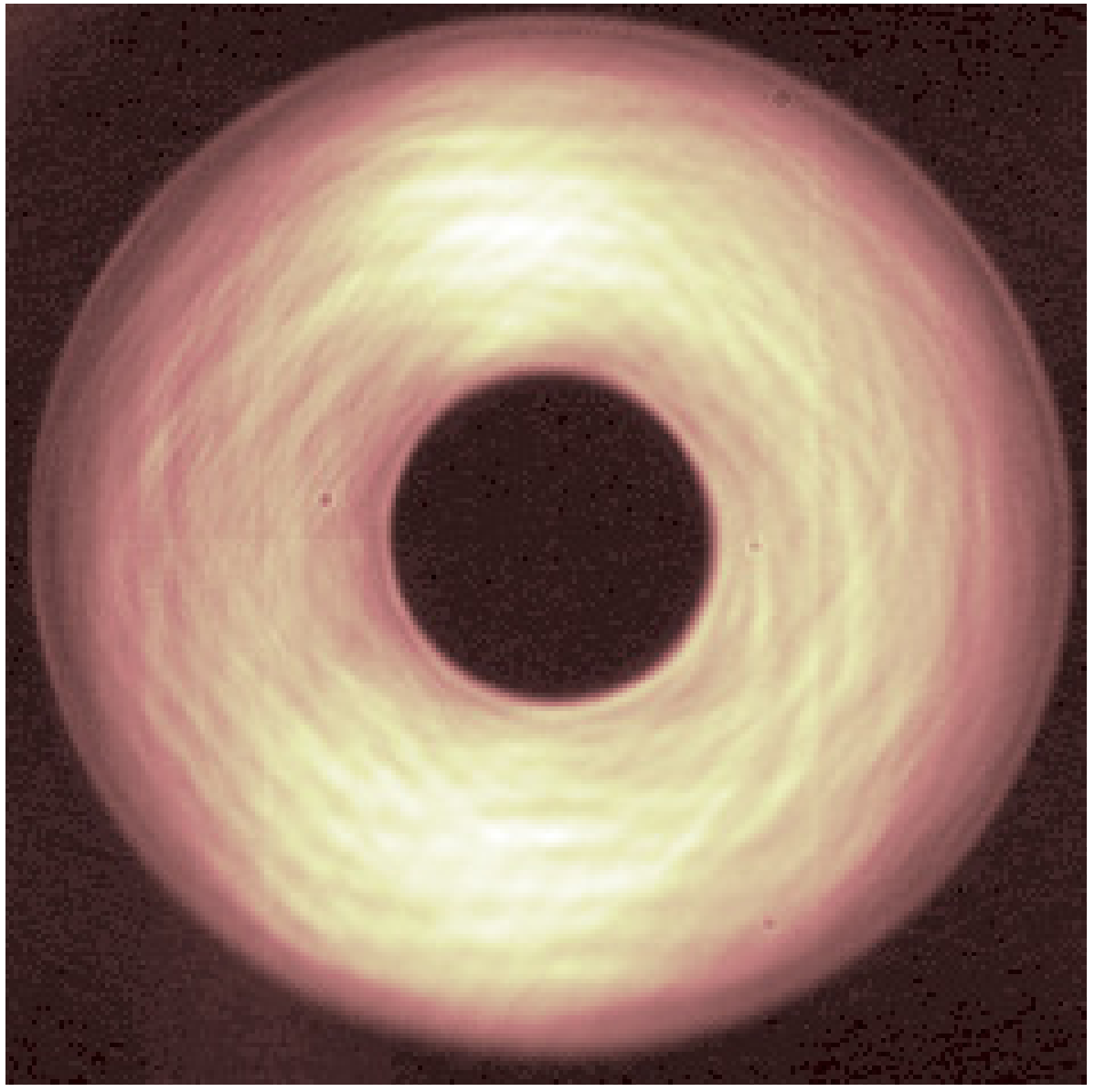}
\\
\begin{subfigure}{\figsize}
\centering
\includegraphics[width=\linewidth]{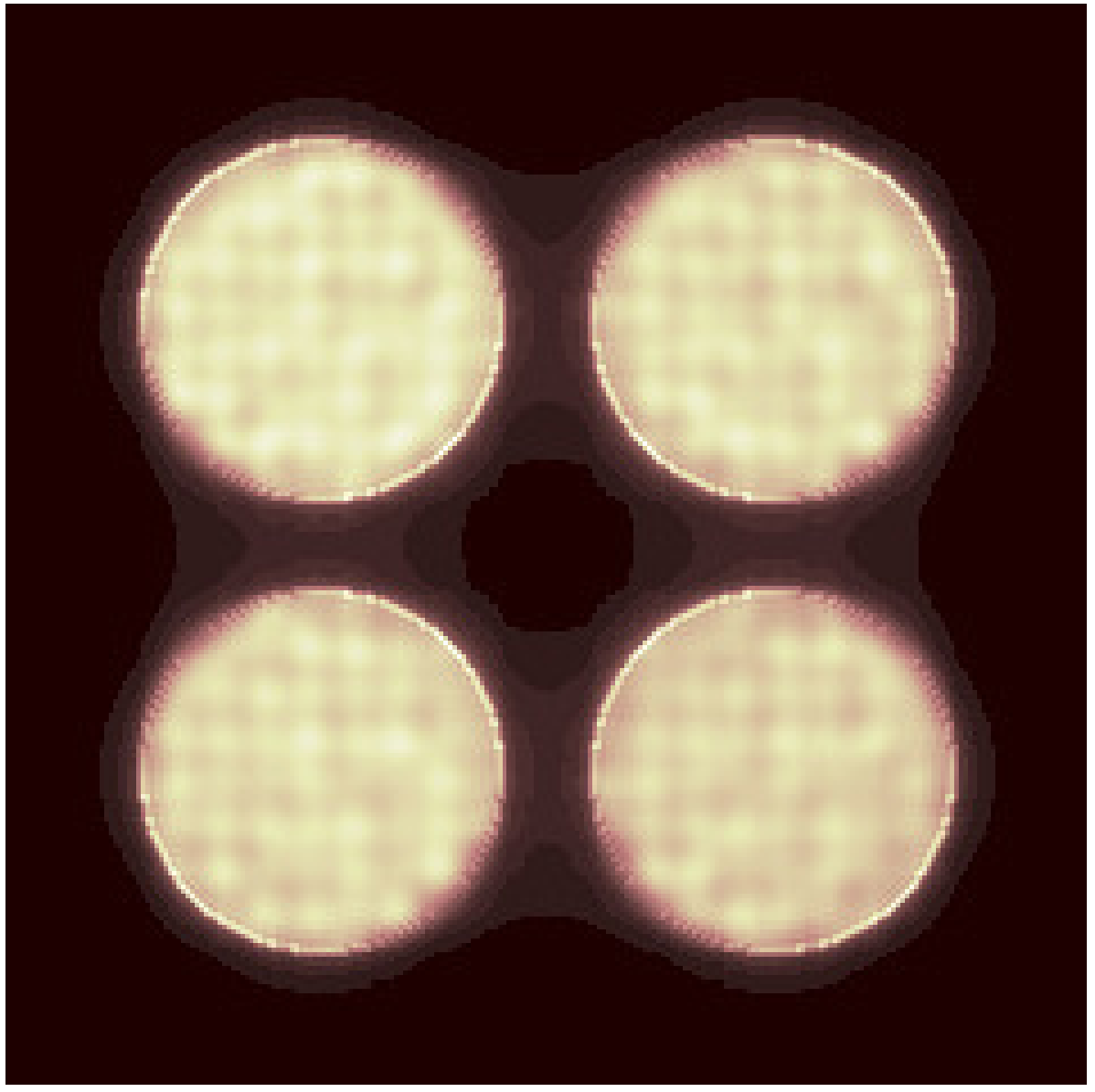}
\caption{}
\label{sub:moda}
\end{subfigure}
\begin{subfigure}{\figsize}
\centering
\includegraphics[width=\linewidth]{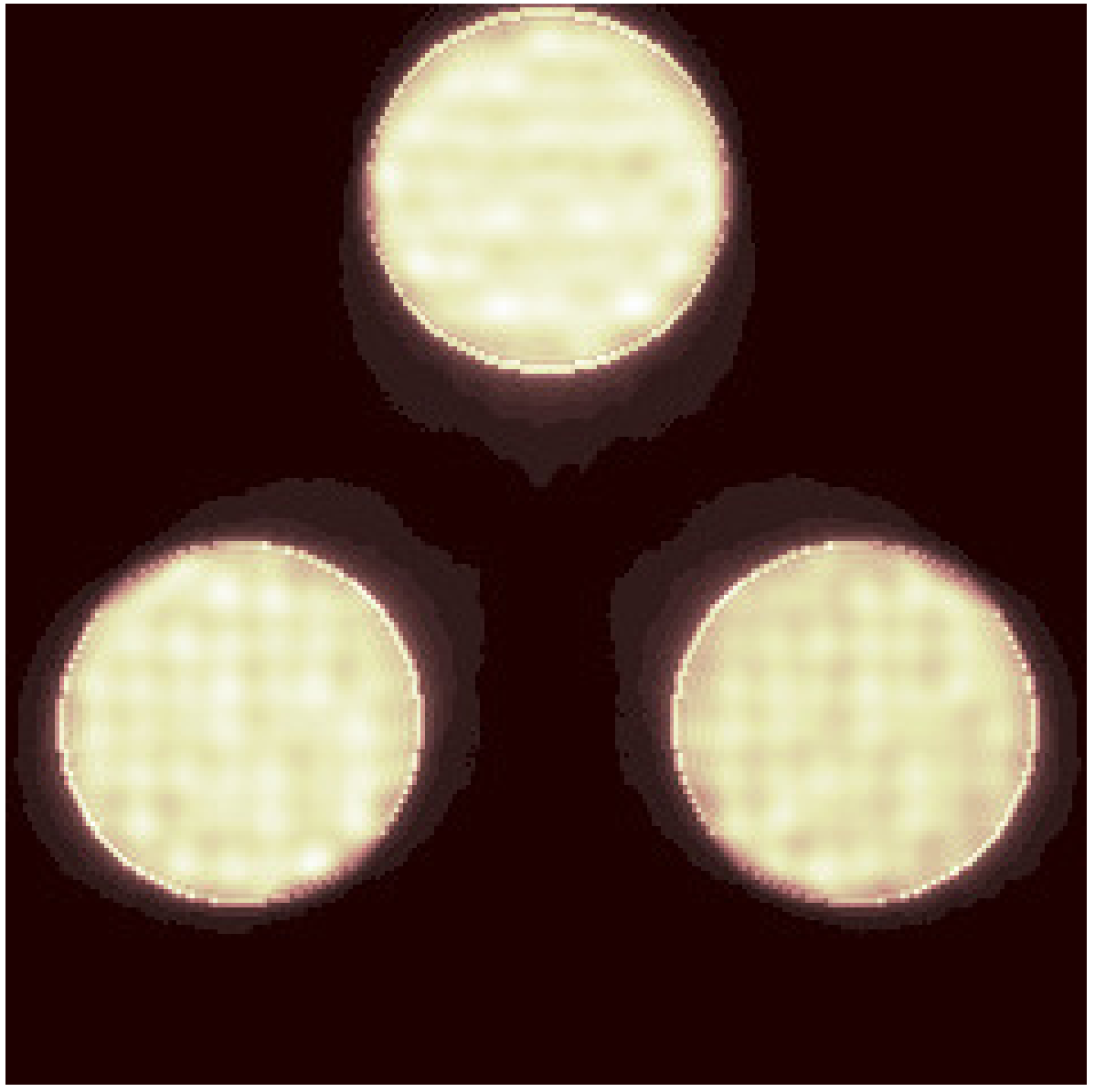}
\caption{}
\label{sub:modb}
\end{subfigure}
\begin{subfigure}{\figsize}
\centering
\includegraphics[width=\linewidth]{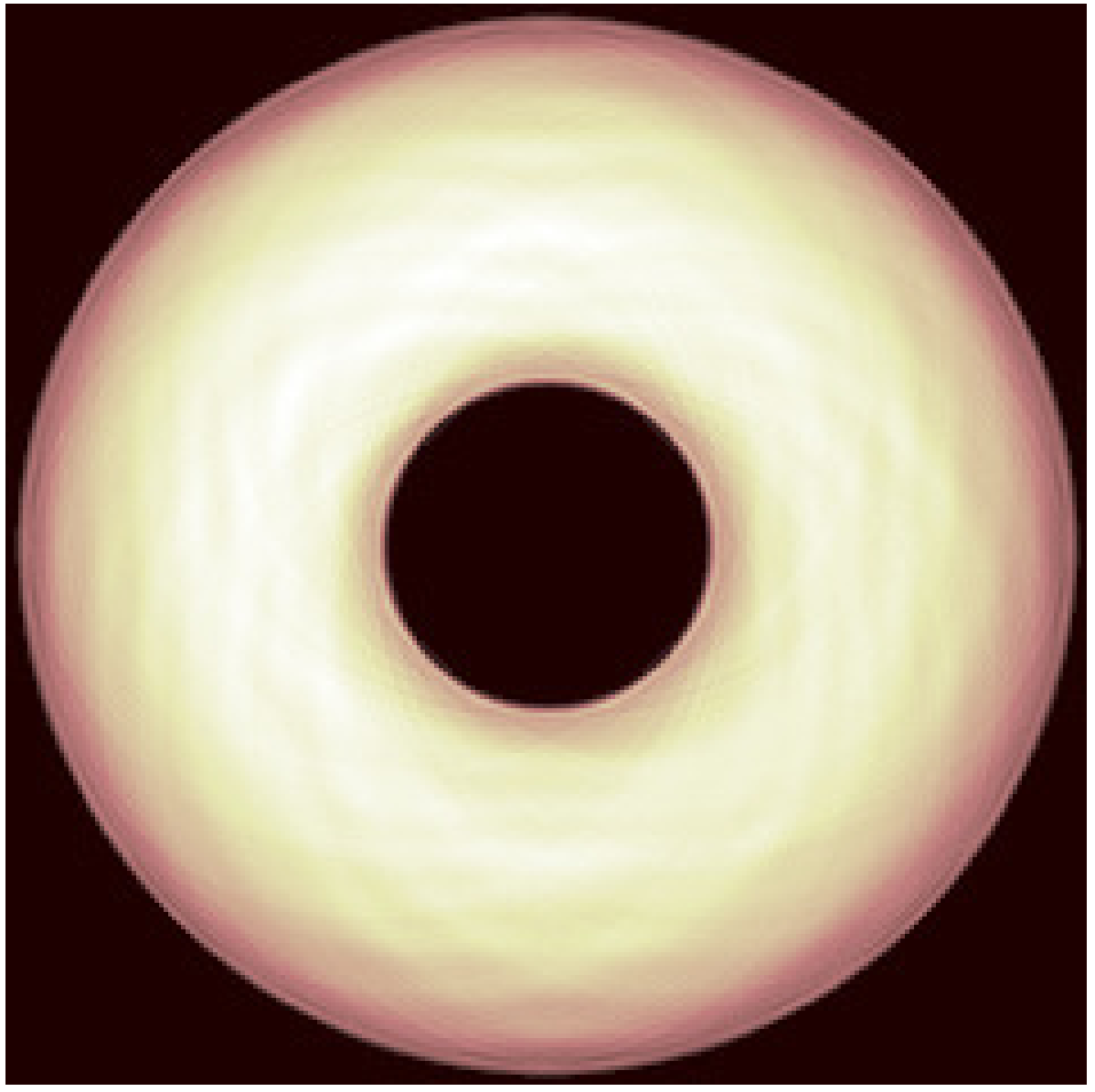}
\caption{}
\label{sub:modc}
\end{subfigure}

\caption{Top: images taken on the LOOPS bench in the pupil plane after the SLM with a modulation of $5\lambda/D$. Bottom: simulation images corresponding to the top cases. The focal masks applied on the SLM are respectively: (a) a 4 faces pyramid, (b) a 3 faces pyramid and (c) an axicone.}
\label{fig:modulatedFWFS}
\end{center}
\end{figure}

Figure~\ref{fig:modulatedFWFS} presents the images obtained with a modulation of $5\lambda/D$ on the bench (top) and in simulation (bottom). Here again the correspondence between simulation and bench images is strong. An elongation of the sub-pupils can be perceived on (a) and (b) from the bench. The same behaviour is present on the simulation image where the contours of the sub-pupils are oriented toward the center of the image. This effect is inherent to the mask that is used, as it divides the PSF into sub-pupils and thus diffracts light in directions given by the mask itself. The greater the modulation radius the lower the effect.

Finally, we calibrated the system by applying Zernike modes on the DM and measuring the resulting intensity distribution on the P-WFS$_2$. We present on Fig.~\ref{fig:calibrationmode} the results obtained when a primary spherical aberration is applied on the DM. The top images are the data acquired on the LOOPS bench and the bottom images data from simulations. The bench data are the image when the DM is set to the given Zernike minus the reference frame when the DM is set to the flat commands. Animations corresponding to the full calibration process are available by following the embedded link in the caption of Fig.~\ref{fig:calibrationmode}.

\renewcommand{\figsize}{0.24\linewidth}
\begin{figure}[tb!]
\begin{center}
\includegraphics[width=\figsize]{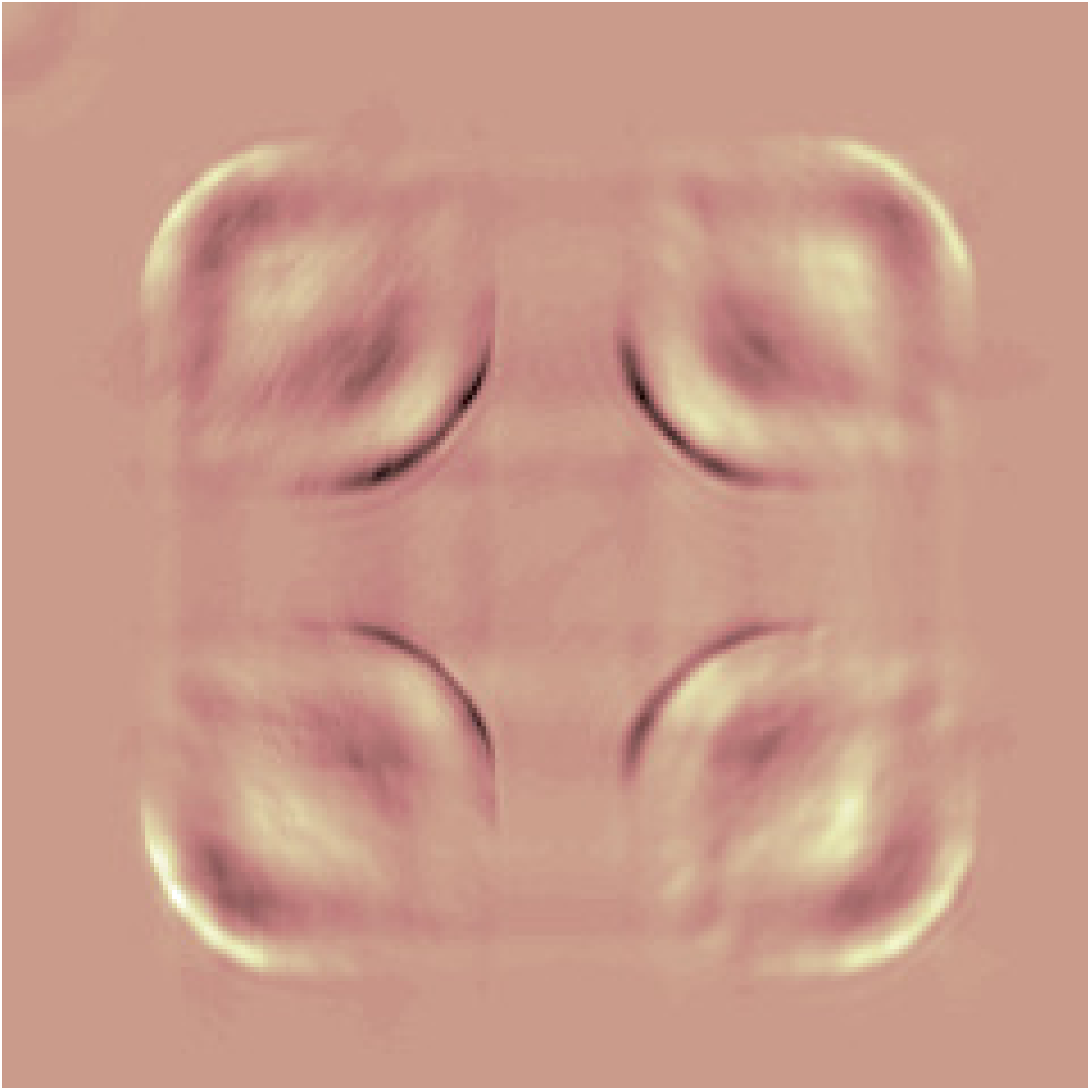}
\includegraphics[width=\figsize]{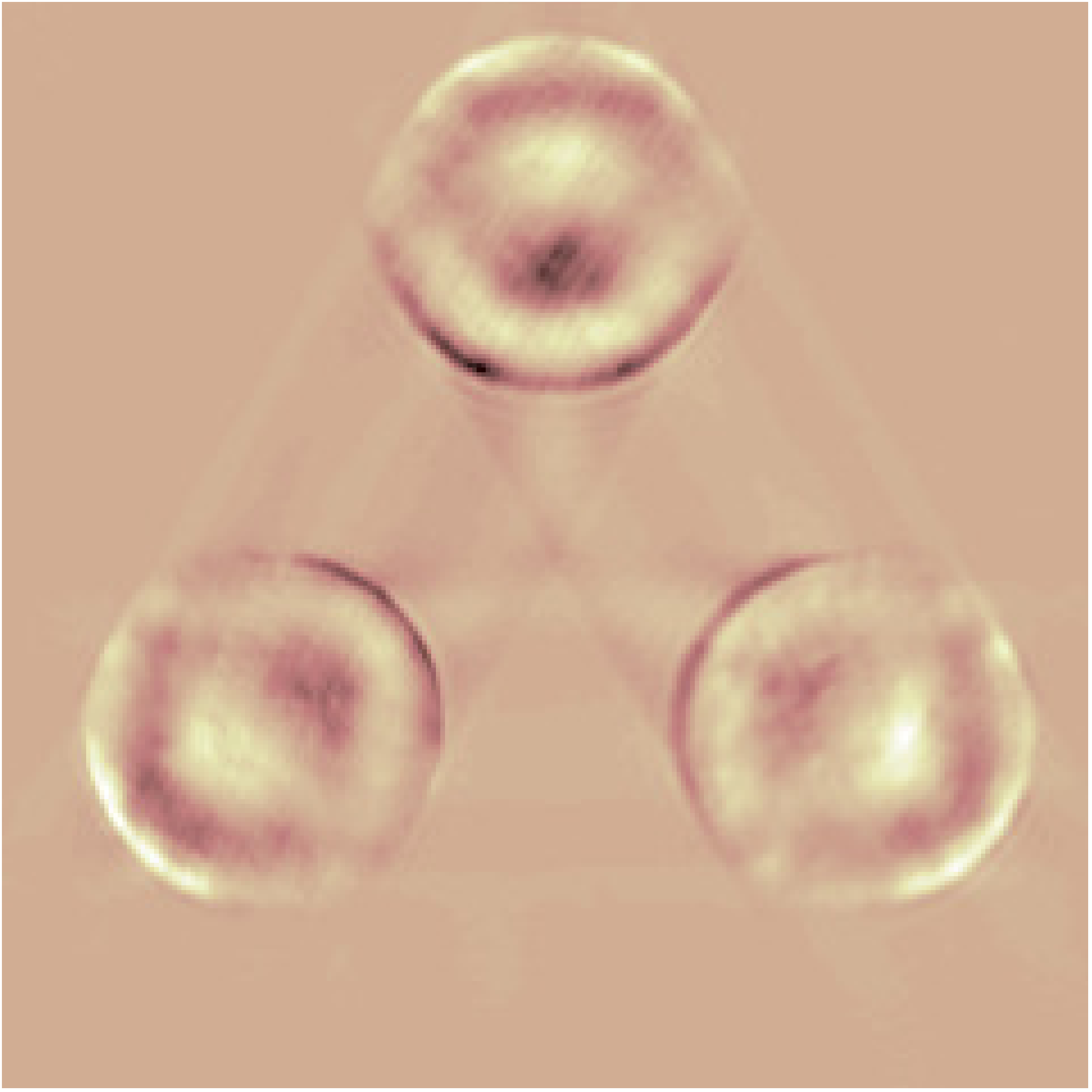}
\includegraphics[width=\figsize]{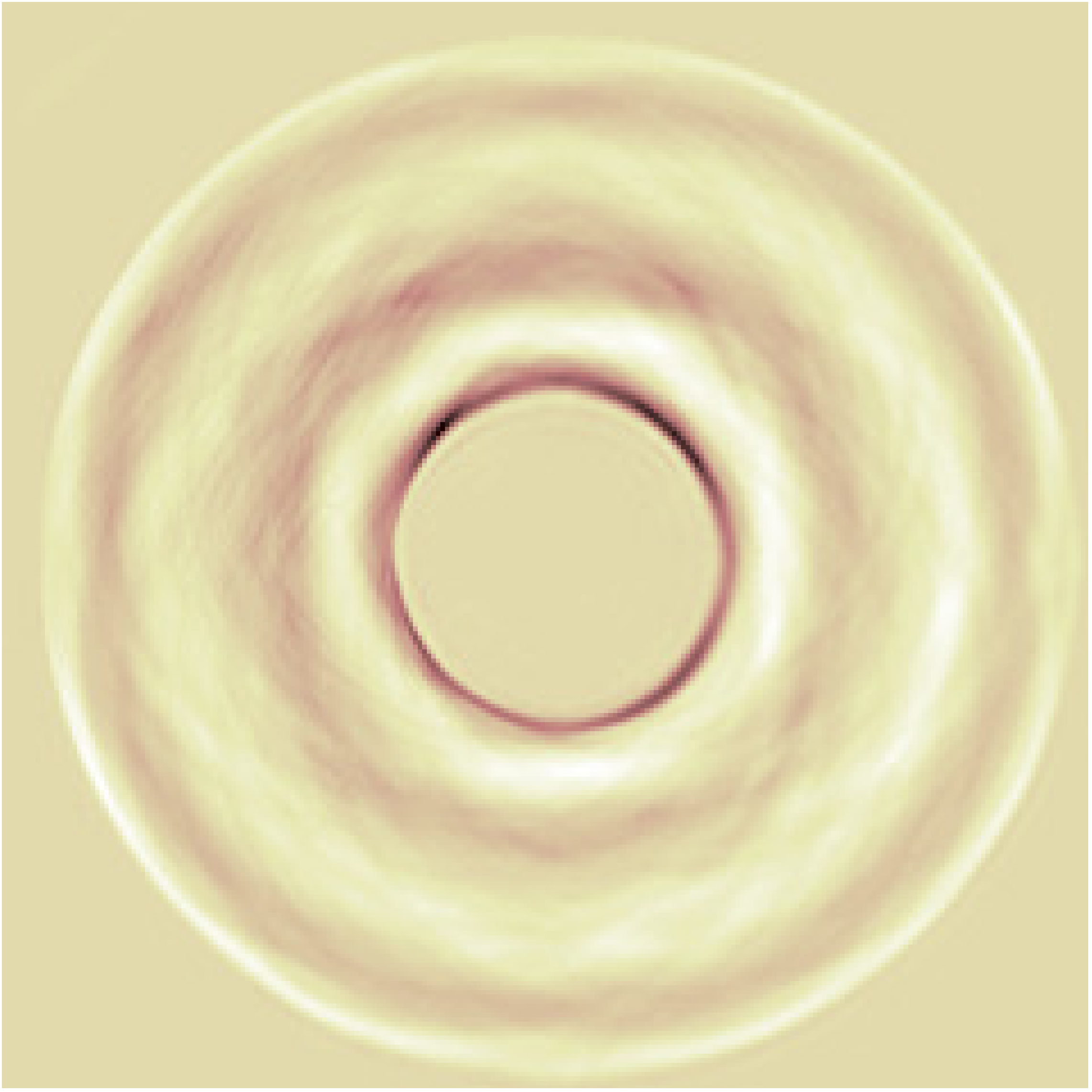}
\includegraphics[width=\figsize]{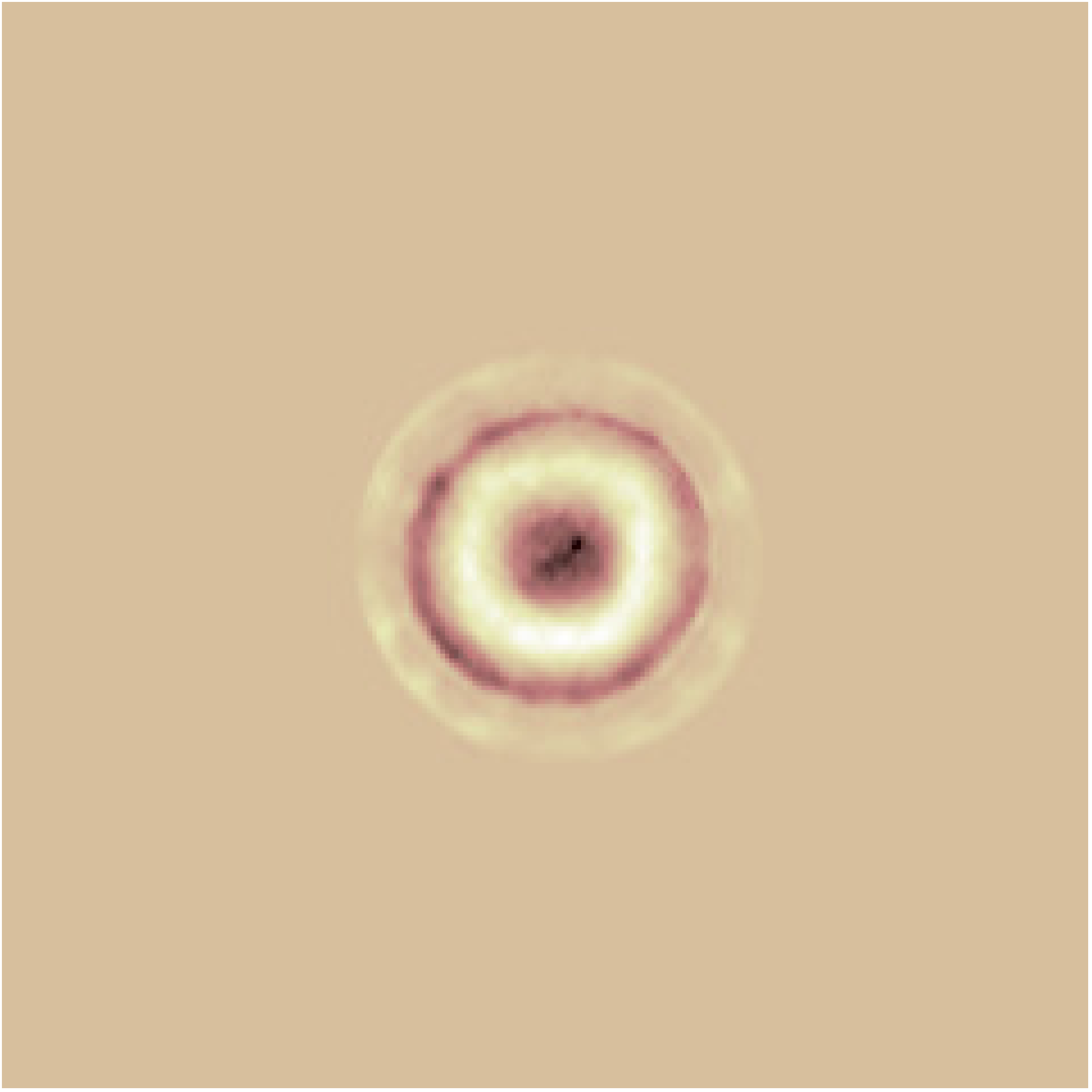}
\\
\begin{subfigure}{\figsize}
\centering
\includegraphics[width=\linewidth]{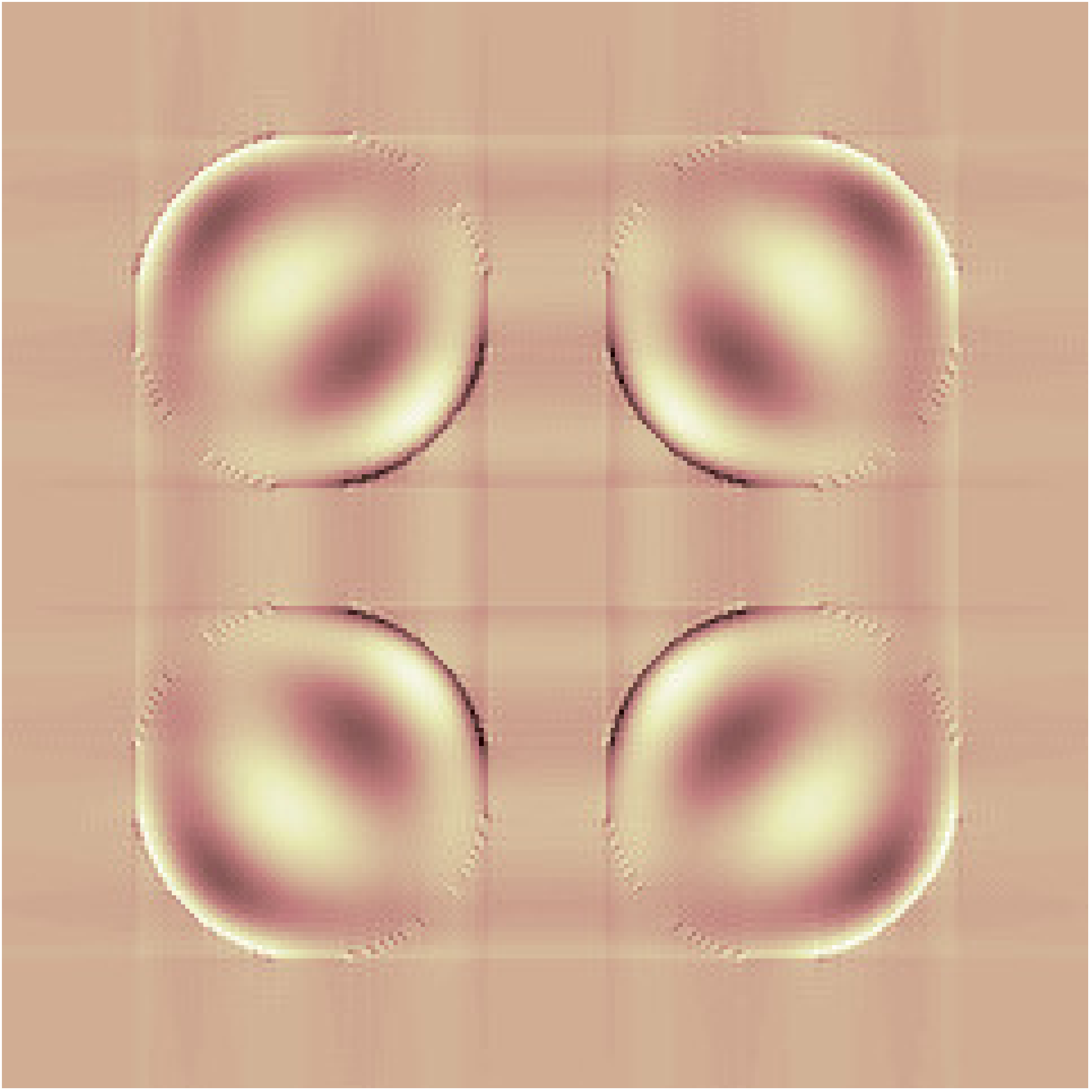}
\caption{}
\label{sub:caliba}
\end{subfigure}
\begin{subfigure}{\figsize}
\centering
\includegraphics[width=\linewidth]{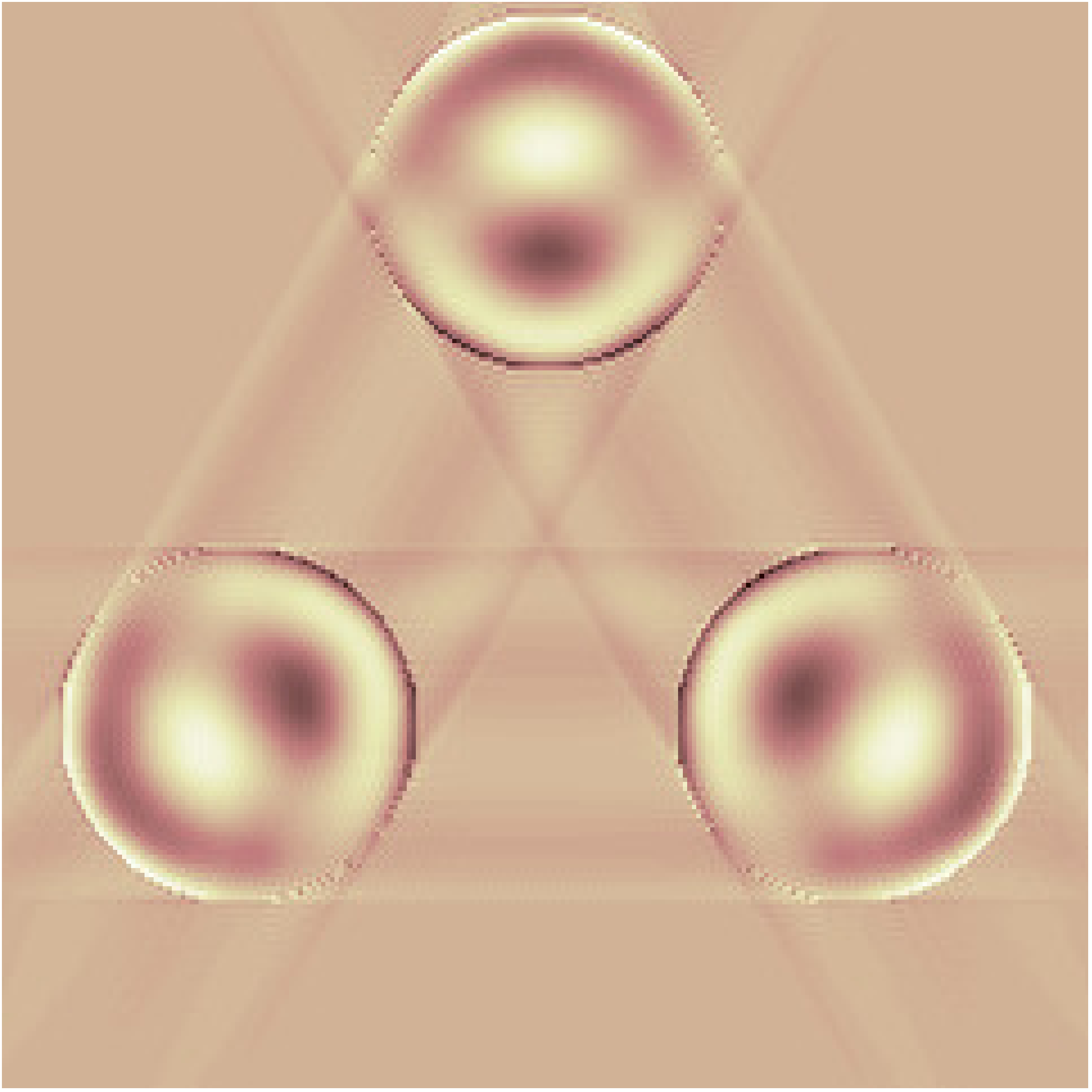}
\caption{}
\label{sub:calibb}
\end{subfigure}
\begin{subfigure}{\figsize}
\centering
\includegraphics[width=\linewidth]{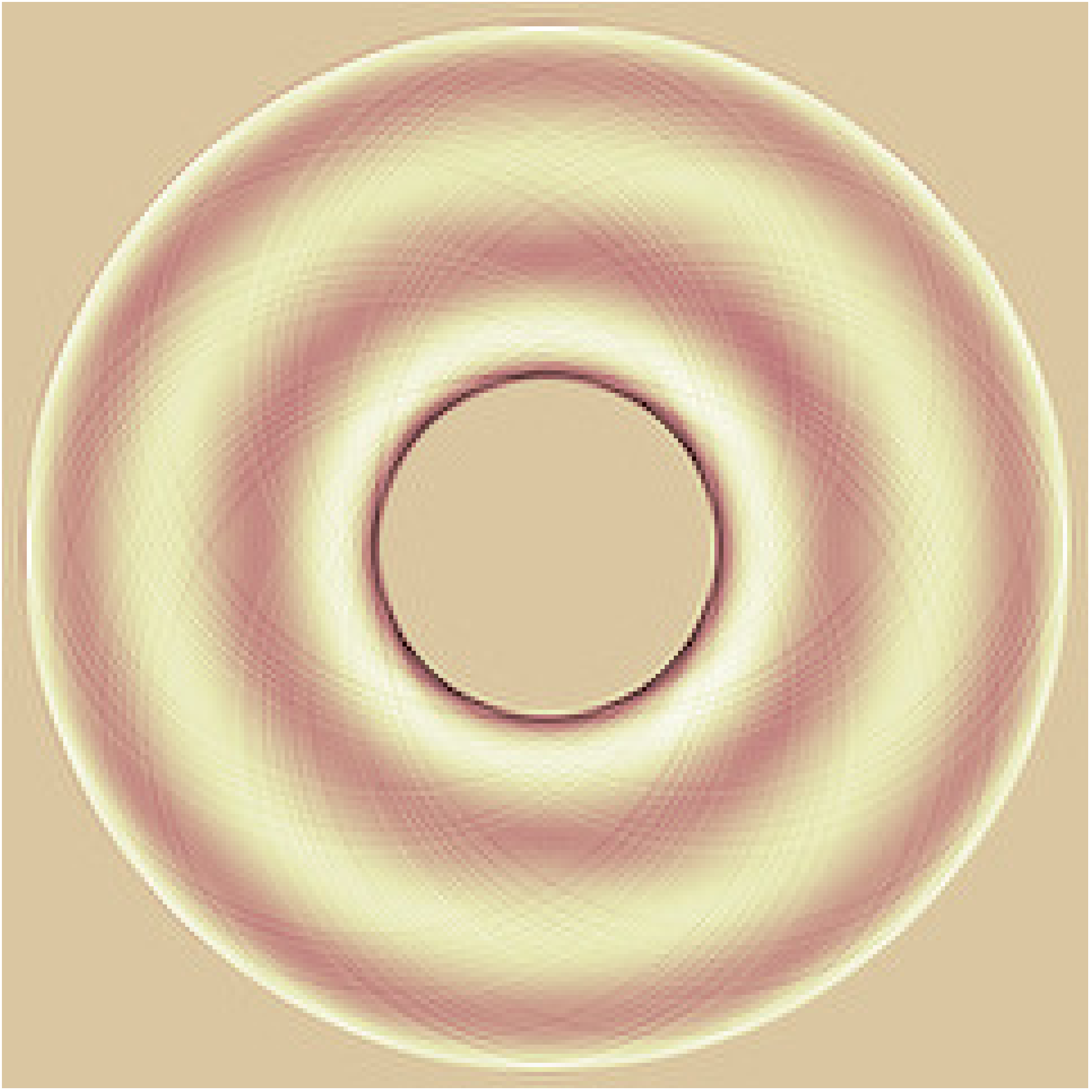}
\caption{}
\label{sub:calibc}
\end{subfigure}
\begin{subfigure}{\figsize}
\centering
\includegraphics[width=\linewidth]{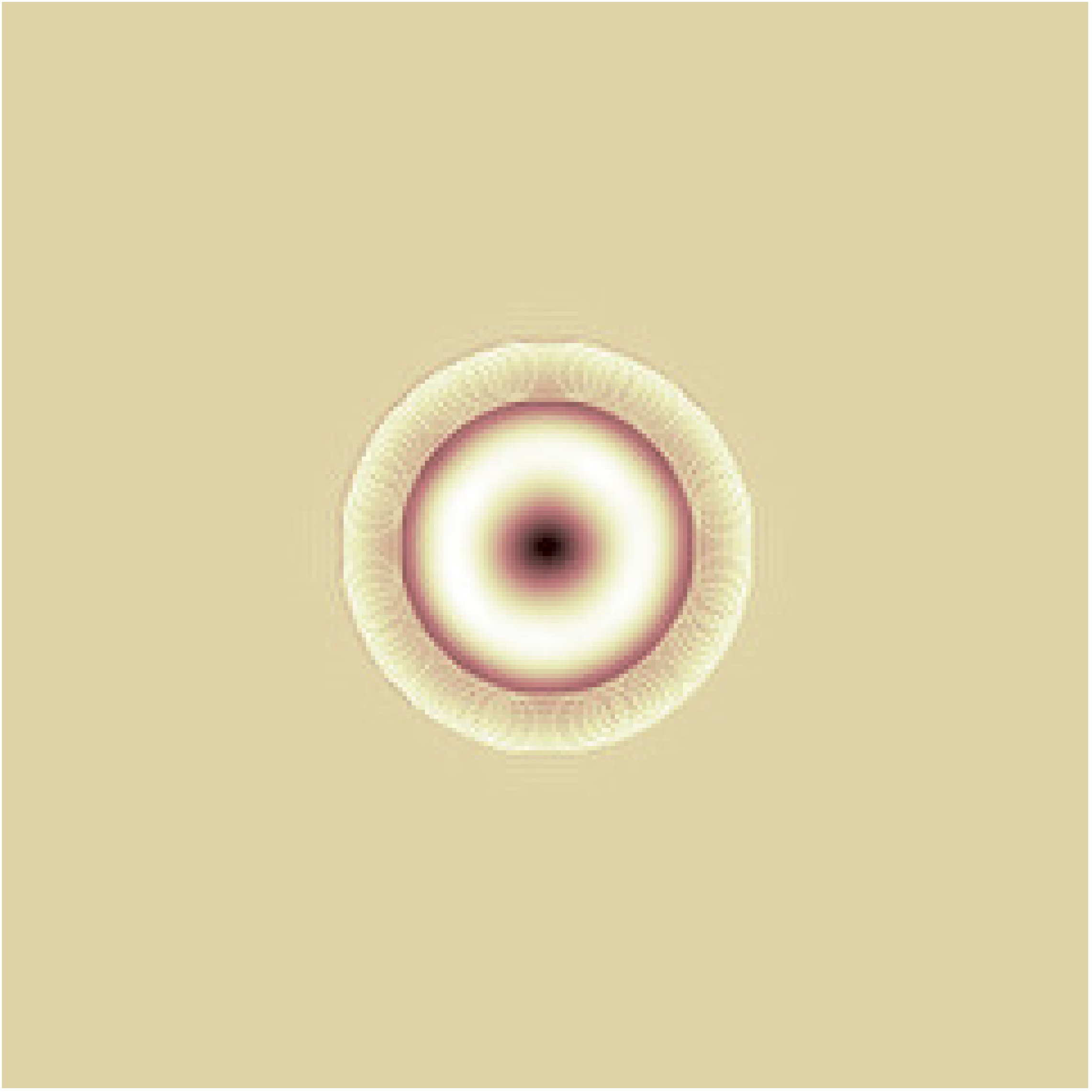}
\caption{}
\label{sub:calibd}
\end{subfigure}

\caption{Top: example of calibration images obtained on the LOOPS bench. Bottom: corresponding images obtained in simulation. In the presented case, we apply a primary spherical aberration on the DM and show the image taken from the camera minus a reference image taken when the DM is flat. The focal masks applied on the SLM are respectively: (a) \href{https://youtu.be/2gQTlA3lNsM}{a 4 faces pyramid}, (b) \href{https://youtu.be/BDRiP2TmdT4}{a 3 faces pyramid}, (c) \href{https://youtu.be/7wih7yPG0yQ}{an axicone} and (d) \href{https://youtu.be/4yfSHBRblR4}{a flattened pyramid}. Clickable links lead to videos showing the full calibration process.}
\label{fig:calibrationmode}
\end{center}
\end{figure}

\begin{figure}[tb!]
\begin{center}
\includegraphics[width=.5\linewidth]{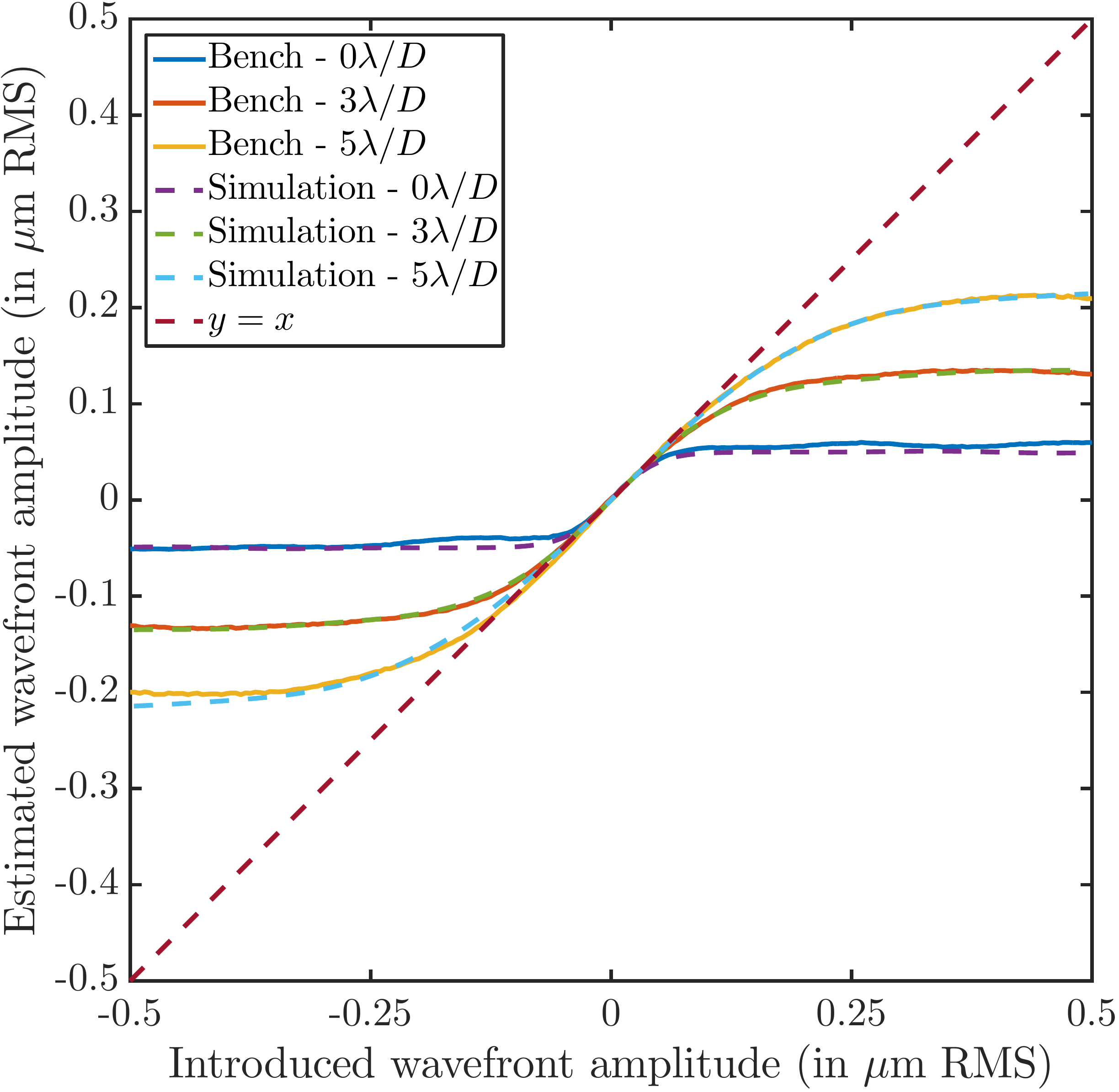}
\caption{Estimated wavefront error versus introduced wavefront error for different modulations. The introduced mode is a primary spherical aberration. The system is inverted with an unfiltered pseudo-inverse of the command matrix obtained with 65 Zernike modes. Solid lines represent the measurements obtained on the bench, dashed lines the simulation data.}
\label{fig:Mode21Linearity}
\end{center}
\end{figure}

As a first probe of wavefront reconstruction, we inverted the calibration matrix acquired for the first 65 Zernike modes and retrieved the amplitude of one introduced mode (primary spherical aberration, referred to as mode \#21 and presented earlier in Fig.\ref{fig:calibrationmode}) for three different modulation radii. The results are presented on Fig.\ref{fig:Mode21Linearity}, the solid lines are the data acquired on the bench and the dashed lines the data produced in simulations. As expected, the greater the modulation radius, the larger the linearity range\cite{}.
This brings confidence that the bench is operating correctly, and that we can reproduce the results accurately in simulation. This leads us to the next section where we present the closed-loop operation results.

%%%%%%%%%%%%%%%%%%%%%%%%%%%%%%%%%%%%%%%%%
% CLOSED LOOP
%%%%%%%%%%%%%%%%%%%%%%%%%%%%%%%%%%%%%%%%%
\section{Closed-loop results with 4PWFS}
\label{sec:cl4PWFS}

We present in this section the results we obtained in closed-loop with a 4 faced pyramid applied on the SLM. The sub-pupils are separated by $1.25 D$ as previously presented in Fig.~\ref{fig:modulatedFWFS}. The DM is controlled using the first 65 Zernike modes. We use the classical slopes-maps approach to handle the wavefront sensor signals.
Before any operation, the static phase aberrations from optics present on the SLM path are minimized by calibrating the unmodulated PWFS and closing the loop on the reference mirror R$_f$1 while setting the reference slopes to zero. We ensure in that way to use a close to optimal PSF at the pyramid apex to avoid optical gain effects that may occur\cite{Korkiakoski08,Deo18}.
We then acquire the calibration matrix with a $5\lambda/D$ modulation and invert it to build the command matrix. The conditioning of the matrix is around 10 and we decide not to filter any modes. The loop gain is set to $0.5$ in order to ensure the stability of the system. We measure the quality of the correction using the Shack-Hartman WFS.

\begin{figure}[tb!]
\begin{center}
\includegraphics[width=.45\linewidth]{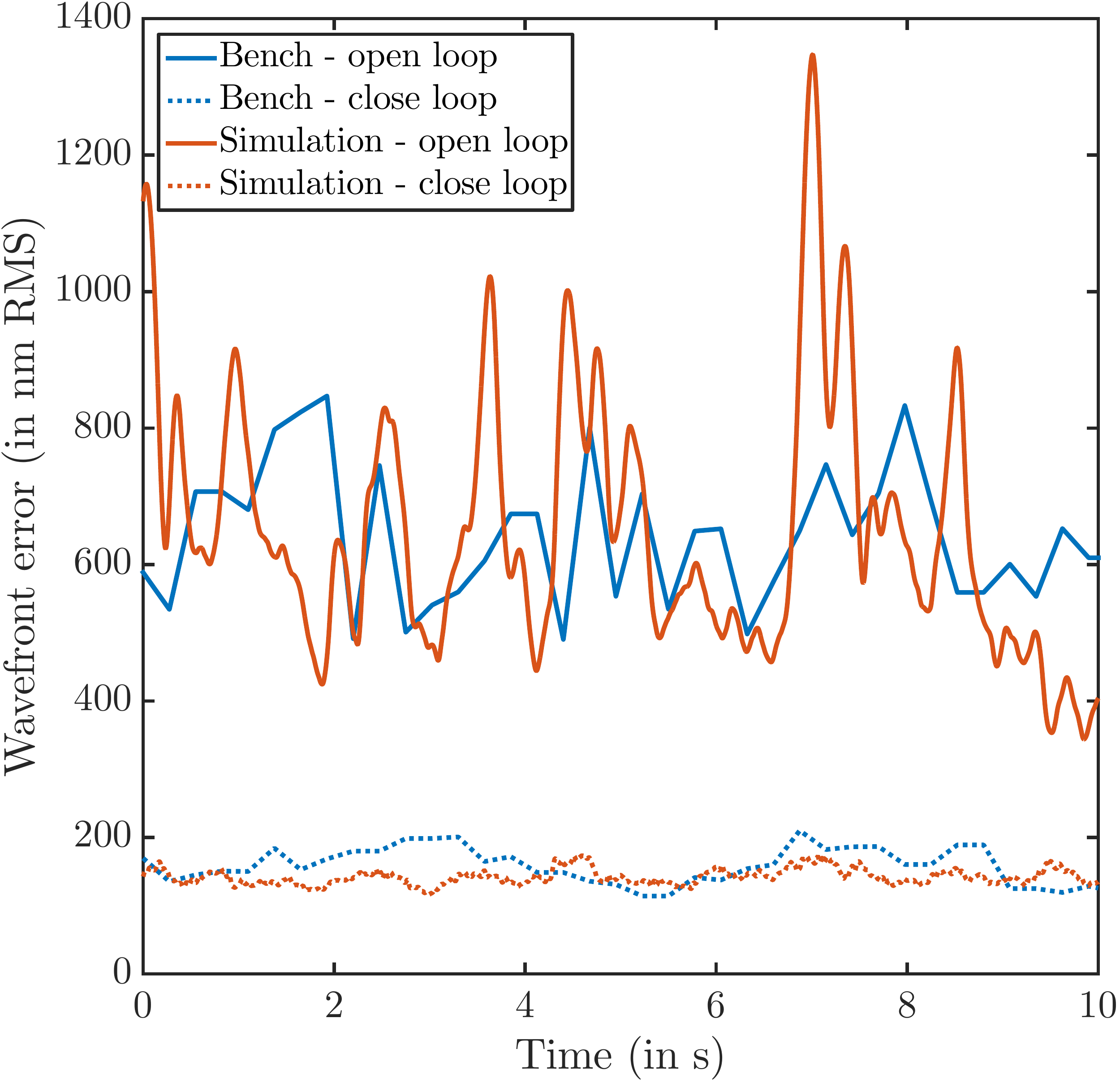}
\hfill
\includegraphics[width=.45\linewidth]{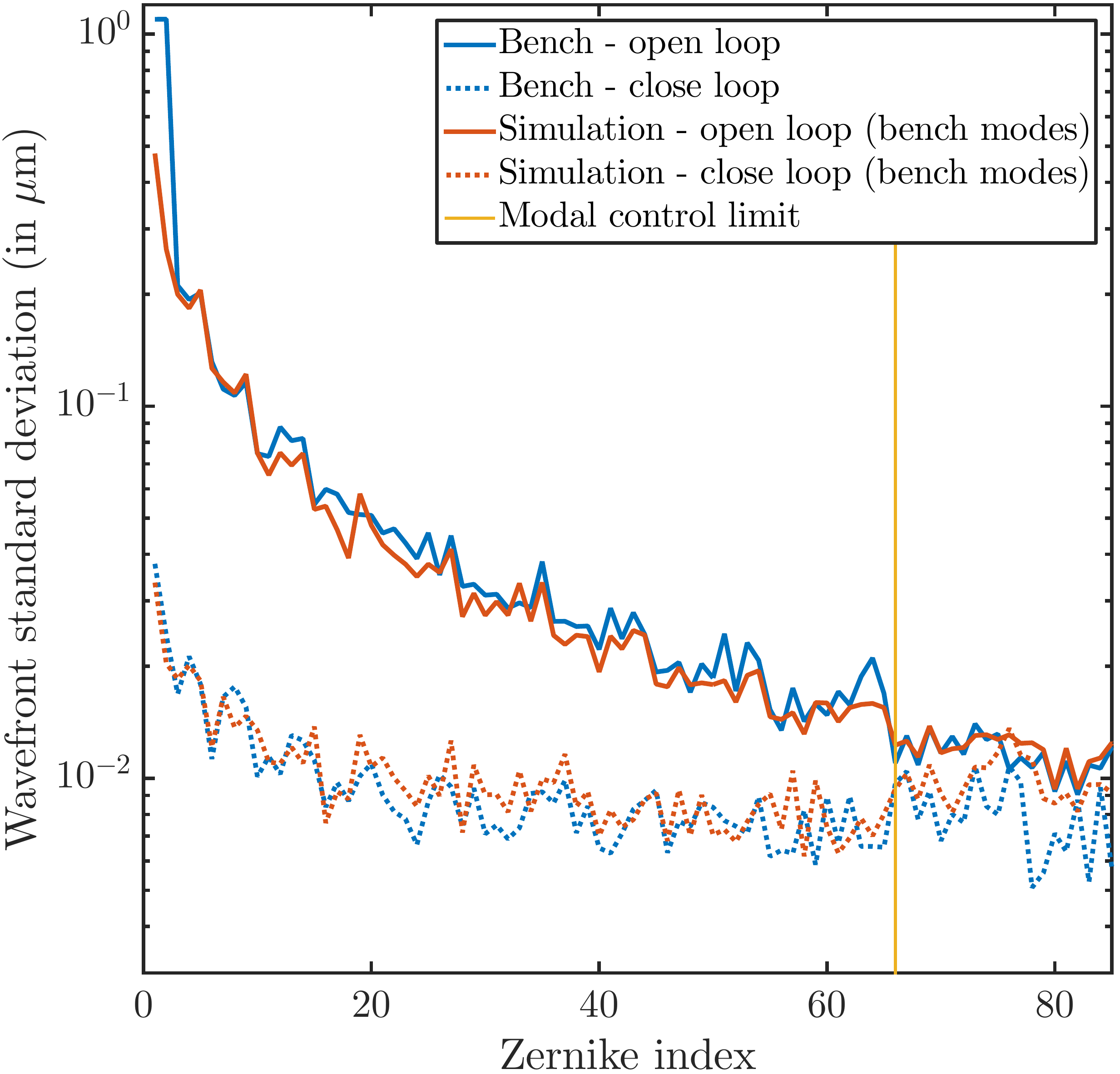}
\caption{Open and closed-loop results for a 4 faced PWFS using slopes-maps and modulated at $5\lambda/D$. Left: RMS wavefront error from 10 seconds open (solid line) and close (dashed line) loop on the bench (blue) and in simulation (red). Right: Standard deviation of the first 130 Zernike modes measured in open loop (solid line) and in closed-loop (dashed line) on the bench (blue) and in simulation (red). The yellow line indicates the number of controlled modes.}
\label{fig:CloseLOOPSTD}
\end{center}
\end{figure}

We present on Fig.\ref{fig:CloseLOOPSTD} (left) a 10 seconds measurement of the residual wavefront error obtained on the bench and reproduced in simulation in open and closed-loop. The residuals when the loop is closed are $\sim150$~nm RMS for both the bench and the simulation.
To go more into details, we present on Fig.\ref{fig:CloseLOOPSTD} (right) the standard deviation of the wavefront error as a function of the Zernike index obtained on the bench and in simulation.
The solid curves represent the open loop scenario. The bench parameters ($r_0$, $L_0$, D) are used in simulation in order to reproduce the turbulent conditions.
Dashed lines present the closed-loop results. We distinguish from modes 1 to 27 a decreasing standard deviation corresponding to the modulation radius. From modes 28 to 65, the standard deviation stay constant as expected for the PWFS after its modulation range. For modes index higher than 66, the system should not correct anymore as we only used the 65 first Zernike modes to control the DM. However, the Zernike modes of the bench are not perfect and present higher frequencies up to what can be produced by the DM. The cutoff is therefore not so pronounced at this location but for frequencies higher than the pitch of the DM, the corrected and uncorrected wavefront superimpose again. We took this effect into account in our simulations by adding higher order modes contributions (from measurements on the SH-WFS) and observed the same behaviour in the cutoff frequency.

We can see a difference in the tip and tilt variance on Fig.\ref{fig:CloseLOOPSTD} (right, modes \#1 and \#2). This additional tip-tilt is introduced by the mount holding the phase plate that produces a wobble of the wavefront. We accounted for this effect on Fig.\ref{fig:CloseLOOPSTD} (left) by subtracting the tip-tilt coming from this wobbling effect.

\begin{figure}[tb!]
\begin{center}
\includegraphics[height=.45\linewidth]{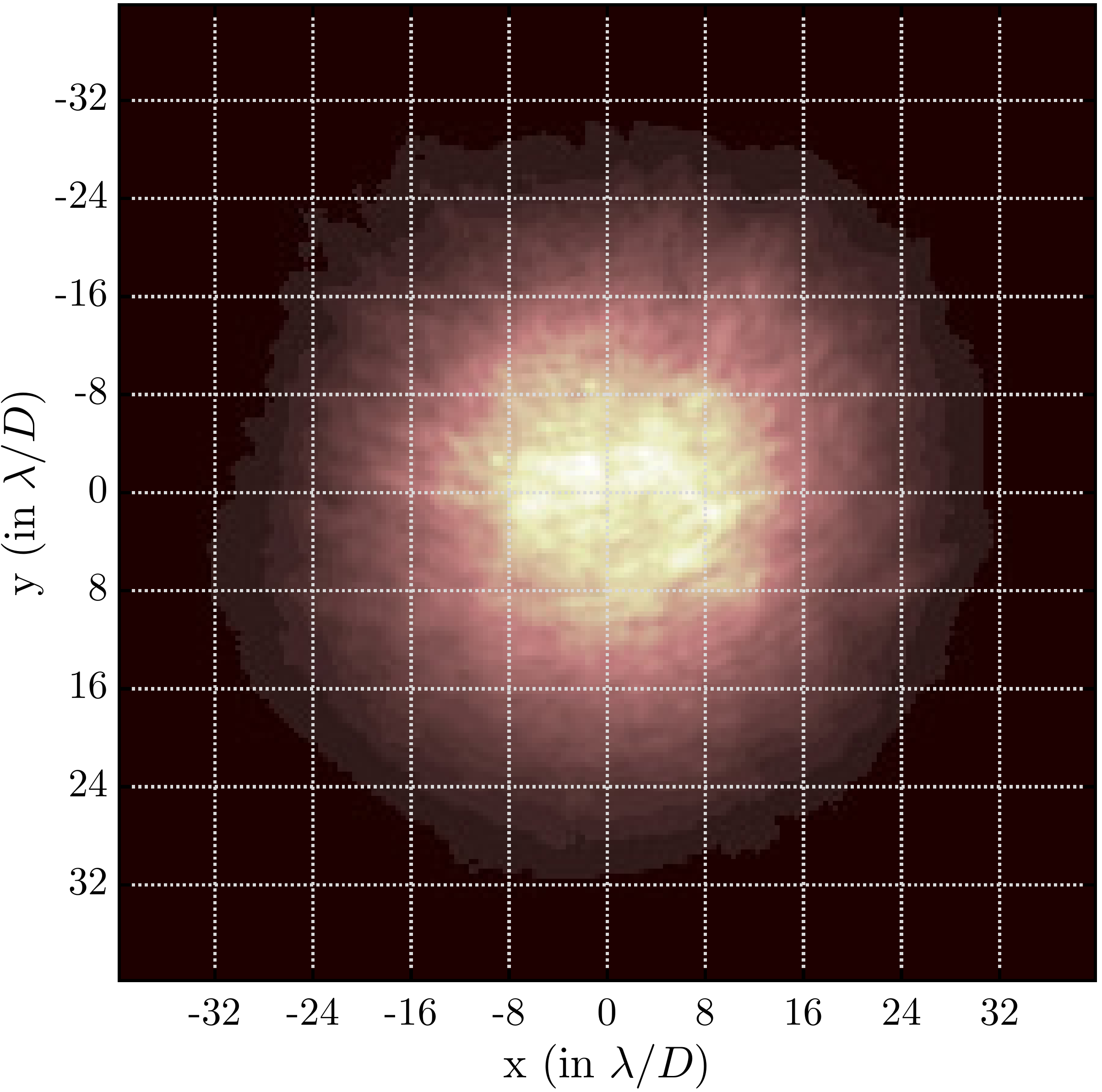}
\hfill
\includegraphics[height=.45\linewidth]{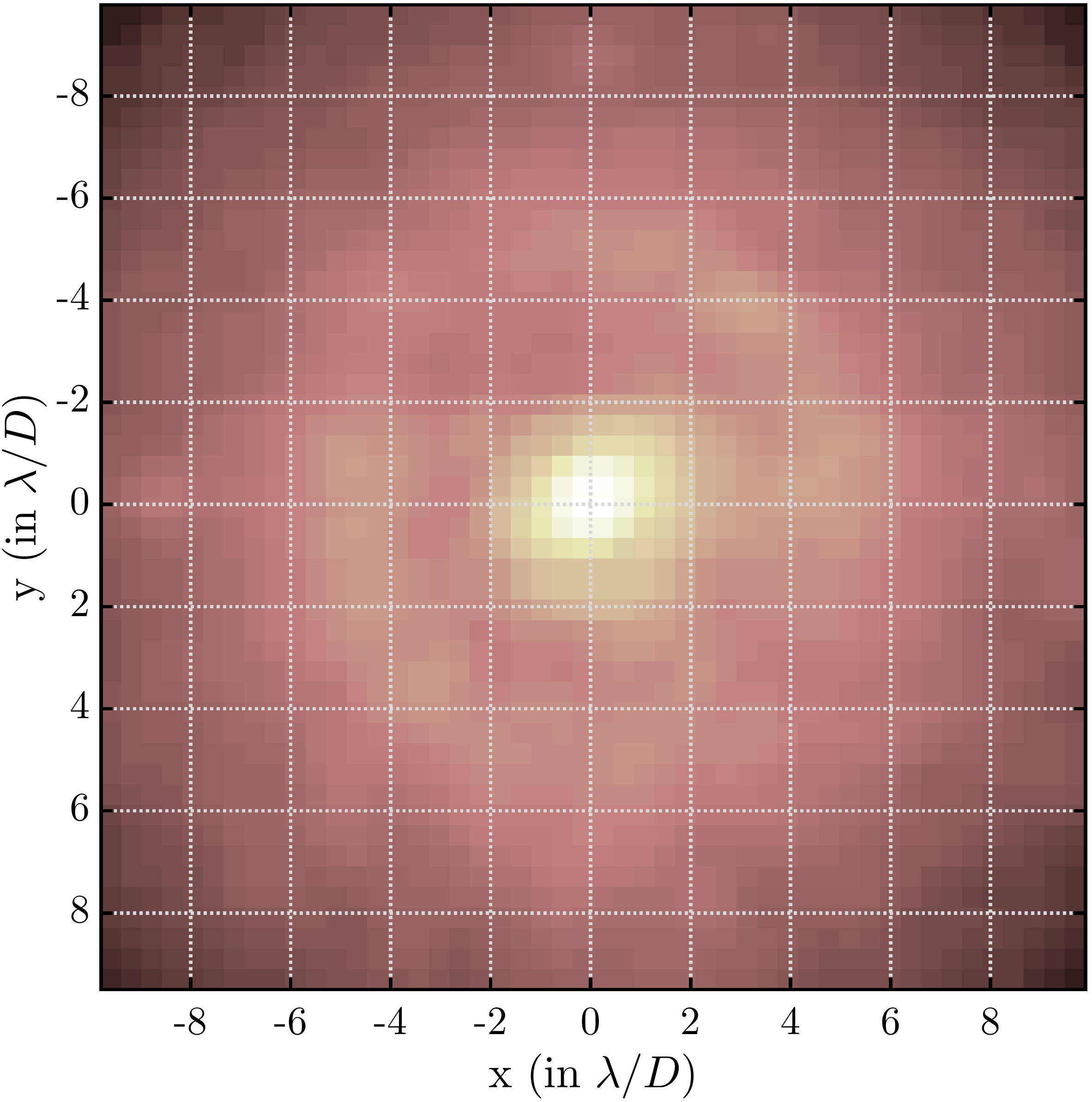}
\caption{Left: open loop long-exposure image with the turbulent phase screen placed in the pupil plane (scale is $40\times40\lambda/D$). Right: log-scale closed-loop long-exposure image with the turbulent phase screen placed in the pupil plane (scale is $10\times10\lambda/D$).}
\label{fig:PSFs}
\end{center}
\end{figure}

The long-exposure PSF are presented on Fig.\ref{fig:PSFs} for the open loop case (left) and the closed-loop case ( right). The scales are different for each image and detailed in the caption. We clearly see on the closed-loop image the correction limits, spanning from $-4\lambda/D$ to $4\lambda/D$, corresponding to the spatial frequency of the actuators spacing in the pupil plane ($8$ inter-actuator spaces comprised in the pupil). The closed-loop PSF is elongated at 45\ensuremath{^\circ}. This is due to the non common path aberrations (NCPA) present between the SLM and the science optical paths because we chose to optimize the PSF at the SLM apex in order to minimize the optical gain effect.

%%%%%%%%%%%%%%%%%%%%%%%%%%%%%%%%%%%%%%%%%
% CONCLUSION
%%%%%%%%%%%%%%%%%%%%%%%%%%%%%%%%%%%%%%%%%
\section{Discussion and conclusion}
\begin{comment}
\noindent
(1) Recall the results obtained from the bench, and especially the first closed-loop operation with different masks on SLM.

\noindent
(2) Recall the bench features and the future applications coming with the next glass pyramid

\noindent
(3) Propose future test we would like to do on the bench like: optical gain compensation, reconstructors, new WFSs ?
\end{comment}

We presented in this paper the latest upgrades made on the LOOPS bench at the Laboratoire d'Astrophysique de Marseille.
We presented in Sec.\ref{sec:LOOPSBench} the updated optical scheme of the bench. It now includes a SLM allowing to create versatile focal phase mask and study their behaviour as wavefront sensors. The bench is controlled with OOMAO and features as well a Shack-Hartman wavefront sensor, a 4 faces glass pyramid wavefront sensor, a $9\times9$ DM and an Hamamatsu Orca Flash camera.
It provides a solid basis for research and development on the Fourier-based wavefront sensors. The comparison between the well mastered Shack-Hartman WFS and new WFS like the 3 faces or flattened pyramids is made easy in this environment.
In a future upgrade of the setup, a SLM will be added in the pupil plane to replace the rotating turbulent phase screen. It will allow the production of adaptable turbulent phase screens with varying parameters ($r_0$, $L_0$, or even different power spectral densities), making the bench capable of reproducing a variety of atmospheric conditions.

We showed in Sec.\ref{sec:SLM} the care we took during the integration of our SLM in a focal plane. Tackling the effects of polarization, diffraction and phase wrapping, we were able to understand in detail the operation of the device, and confirmed the results comparing the images obtained in simulation and on the bench. Shifting the signal out of the $0^{th}$ diffraction order, we enable the possibility of creating WFS with overlapping subpupils like the flattened pyramid or Zernike WFSs.

In Sec.\ref{sec:firstimages} we presented the first images obtained with different WFS with and without modulation as well as phase estimation with a SLM-made 4 faced PWFS. The data acquired from the bench fit well with the data we obtained by mean of numerical simulations. This ensures the SLM is behaving as we want and can reproduce testbench data in simulations. We have now a fully operational testbed to evaluate and characterize new Fourier-based WFS. We presented the first calibration results at the end of Sec.\ref{sec:firstimages} and we are now investigating the performances of new WFS.

Finally, we presented in Sec.\ref{sec:cl4PWFS} the closed-loop result we obtained from a SLM-made 4-faced PWFS. We measured the residual wavefront error in both open and closed-loop situation and were able to reproduce with great accuracy these behaviours in simulation. The LOOPS bench is now ready to perform work on characterizing and developing new WFS and new wavefront sensing techniques.

This work represents a major step and is intended for the next coming papers. It will include linearity and sensitivity measurements as well as bootstrapping and closed-loop analysis for various Fourier-based WFSs. The points of peculiar interest are (1) the comparison between the 4 and 3 faces pyramids using a slopes map approach and (2) the comparison between the slopes map and full frame approaches on the 4 faces pyramid.

\section*{Funding}
P. Janin-Potiron is grateful to the French Aerospace Lab (ONERA) for supporting his postdoctoral fellowship.
This work has been partially supported by the LABEX FOCUS (grant DIR-PDC-2016-TF) and the VASCO research program at ONERA. It also benefited from the support of the project WOLF ANR-18-CE31-0018 of the French National Research Agency (ANR) and the A*MIDEX project (no. ANR-11- IDEX-0001- 02) funded by the ”Investissements d'Avenir” French Government program, managed by the French National Research Agency (ANR) and the Action Spécifique Haute R\'esolution Angulaire (ASHRA) of CNRS/INSU co-funded by CNES.

% Bibliography
\bibliographystyle{spiejour}
\bibliography{closedloop}

\end{spacing}
\end{document}